\documentclass[pra,preprint,showpacs]{revtex4}
\usepackage[centertags]{amsmath}
\usepackage{amsfonts}
\usepackage{amssymb}
\usepackage{amsthm} 
\usepackage{newlfont}
\usepackage{epsfig}
\usepackage{amscd}
\usepackage{graphicx}
\usepackage{epsfig}
\usepackage{amscd}

\newcommand{\beq}{\begin{equation}}
\newcommand{\eeq}{\end{equation}}
\newcommand{\ba}{\begin{array}}
\newcommand{\ea}{\end{array}}
\newcommand{\bea}{\begin{eqnarray}}
\newcommand{\eea}{\end{eqnarray}}
\begin{document}

\begin{center}
{\large \sc \bf { Remote control of quantum correlations in two-qubit receiver via three-qubit sender
}}

\vskip 15pt

{\large 
S.I.Doronin and A.I.~Zenchuk 
}

\vskip 8pt

{\it Institute of Problems of Chemical Physics, RAS,
Chernogolovka, Moscow reg., 142432, Russia},\\

\end{center}


\begin{abstract}
We study the problem of remote control of quantum correlations (discord) 
in a sub-system of two qubits (receiver) via the parameters of the initial state  of another 
sub-system of three qubits
(sender) connected with the receiver by  the inhomogeneous  spin-1/2 chain.
We propose two parameters characterizing the creatable correlations. The first one is  
 the discord between the  receiver and 
the rest of  spin-1/2 chain, it concerns the mutual correlations between these two subsystems. 
The second parameter 
is  the discord between the two nodes of the receiver and 
describes the  correlations  inside of the receiver. We study the dependence of these two discords on the 
inhomogeneity degree of  spin chain.
 \end{abstract}

\maketitle

\section{Introduction}
\label{Sec:Introduction}

The problem of controllable remote state creation considered in set of papers \cite{PBGWK2,PBGWK,DLMRKBPVZBW,XLYG,PSB} 
initiates  the 
problem of creation the states with desirable quantum correlations in a
 receiver. In particular, the  entanglement between the remote qubits is studied in  \cite{BBVB,CS}, 
 different method of creation of quantum correlations are considered in 
 \cite{DSC,LBAW,NLLZ,ZC,SXSZDWHCKW,RDL}.
 
We shall recall that the problem of remote state control has rather long 
history starting  with the quantum echo \cite{FBE}, which can be referred to as 
the long distance quantum state transfer. The problem of quantum state transfer itself was formulated in 
\cite{Bose}. But even earlier the problem of quantum teleportation was stated \cite{BBCJPW}. 
It is worthwhile to give a brief comparison 
of such  closely related branches  of quantum communication as 
quantum teleportation \cite{BBCJPW,BPMEWZ,BBMHP}, quantum state transfer \cite{Bose,CDEL,ACDE,KS,GKMT} 
and remote quantum state creations \cite{PBGWK2,BDSSBW,BHLSW,G,PBGWK}. 

The teleportation of the unknown state differs from the two others by the presence of the additional  classical communication channel. 
However, in some sense, this channel is implicitly implemented into the  interaction Hamiltonian 
governing the dynamics of the communication line in the process of state transfer and state creation. 
 The simple analogy can be observed 
in the perfect state transfer,  when the unknown  sender's state 
moves to the receiver. So, the  classical channel as an additional part of the 
''communication line'' is not needed.
Next, the high probability state transfer \cite{KZ_2008,SAOZ,BK,C,QWL,B,JKSS,SO,BB,SJBB} 
was proposed, which  is much simpler realizable in comparison with  the perfect state transfer. 
Besides, instead of transferring the sender's state itself,  we may try
to create another state-of-interest directly related with the sender's state (but different from it) 
\cite{PBGWK2,BDSSBW,BHLSW,G,PBGWK}. 
The creation of such states via the spin chain is the subject of ref.\cite{Z_2014}, where this 
idea was formulated for the case of mixed sender's state and short chains. 
The state creation controlled by the   pure sender's state 
with one-spin excitation  was studied in \cite{BZ_2015}, the similar problem with the physically motivated initial state 
is considered in \cite{FKZ_LANL}. We shall also remark that our algorithm of 
the quantum state creation develops ideas  of the quantum information transfer \cite{YBB,Z_2012,PS} which is an 
alternative process to the quantum state transfer.

In this paper we consider the remote state-creation in terms of the quantum correlations described 
by the quantum discord \cite{HV,OZ,Zurek},
which was introduced after the quantum entanglement \cite{Wootters,HW,P,AFOV,HHHH}. As quantum correlation parameters,
we use the  discord between the receiver and the rest of spin chain (external correlations) and the discord between
the nodes of the receiver
(inside correlations). We show their mutual relation and study the map from the control-parameter domain into the 
two-dimensional space of the mentioned above discords.

The paper is organized as follows.
In Sec.\ref{Section:model} we describe our model of communication line 
including the Hamiltonian and the initial state. Quantum correlations at the receiver side are considered in Sec.\ref{Section:Q}. 
The time optimization of the state creation is performed in Sec.\ref{Section:opt}.
Results of the numerical simulation of the creatable correlations  in long chains with different inhomogeneity degrees 
are represented in Sec.\ref{Section:num}.
Conclusions are given in Sec.\ref{Section:conclusion}.

\section{Model of communication line}
\label{Section:model}
 
\begin{figure*}
   \epsfig{file=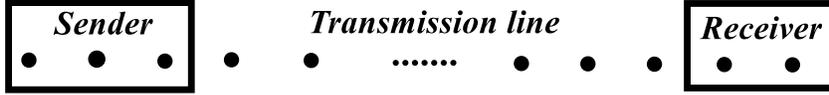,
  scale=0.45
   ,angle=0
}
\caption{The communication line with three-node sender and two-node receiver
} 
  \label{Fig:comm} 
\end{figure*}

The communication line considered in our paper is shown in Fig.\ref{Fig:comm}. It consists of 
the three-node sender (the first three nodes of the chain), the two-node receiver (the two last nodes of the chain)
and the transmission line 
connecting them.

 \subsection{Interaction Hamiltonian}
 We consider the evolution governed 
by the nearest neighbor XY-Hamiltonian 
 \begin{eqnarray}\label{XY}
H=  \sum_{i=1}^{N-1}D_i (I_{ix} I_{(i+1)x} + I_{iy} I_{(i+1)y})
,
\end{eqnarray}
where  $D_i$ are  the coupling constants between the 
nearest neighbors, $I_{j\alpha}$ ($j=1,\dots,N$, $\alpha=x,y,z$) is the 
$j$th spin projection on the $\alpha$-axis.
In our model we use the dimensionless time and 
 the following coupling constants:
\begin{eqnarray}\label{Dnonhom}
D_i=\frac{\sqrt{N-1}\cos(\phi \pi) + \sin(\phi \pi) \sqrt{i(N-i)}}{\sqrt{N-1}(\cos(\phi\pi) + \sin(\phi\pi))},
\;\; 0\le \phi \le \frac{1}{2}.
\end{eqnarray}
The parameter $\phi$ in eq.(\ref{Dnonhom}) indicates the deviation of our chain from the Ekert one and is referred to as the 
inhomogeneity parameter.
Thus
\begin{eqnarray}\label{hom}
D_i|_{\phi=0}=1,\;\;\; \;\;\;{\mbox{homogeneous chain}},\\\label{Ekert}
D_i|_{\phi=\frac{1}{2}}=\sqrt{\frac{i(N-i)}{N-1}},\;\;\;\;\;\;{\mbox{Ekert chain}}.
\end{eqnarray}
Obviously, Hamiltonian (\ref{XY}) commutes with the $z$-projection of the total spin momentum,
$[H,I_z]=0$.
This allows us to significantly simplify the numerical simulations reducing the dimensionality of the Hilbert 
space in which the spin-dynamics is described. So, working with the one-spin excitation, we 
use only $N$-dimensional subspace (of the whole $2^N$-dimensional Hilbert space of the $N$-node spin system) spanned by 
the following vectors:
\begin{eqnarray}
|n\rangle \equiv |\underbrace{0\dots 0}_{n-1}1\underbrace{0\dots 0}_{N-n}\rangle, \;\;n=1,\dots,N.
\end{eqnarray}

\subsection{Initial state of spin chain}

We consider the pure one-excitation initial state of the three-node sender of the  following general 
 form:
 \begin{eqnarray}\label{sins}
&&
|\Psi_0\rangle=\sum_{i=1}^3 a_i|i\rangle ,\\\label{sinsconstr}
&&
\sum_{i=1}^3 |a_i|^2 =1,
 \end{eqnarray}
 where $a_i$ ($i=1,2,3$) are arbitrary  parameters with  constraint (\ref{sinsconstr}). 
 Unlike the initial states considered in \cite{BZ_2015}, our initial state  does
 not involve the ground state $|0\rangle$. According to  the Schr\"odinger equation,
the evolution  of the pure initial state  $|\Psi_0\rangle$ 
 reads:
\begin{eqnarray}\label{ev0}
|\Psi(t)\rangle = e^{-i H t} | \Psi_0\rangle.
\end{eqnarray}
Hereafter we use 
  the following parameterization of the sender's initial state (\ref{sins}) 
  satisfying constraint (\ref{sinsconstr}):
 \begin{eqnarray}
 \label{aalpha}
 a_1=\cos\frac{\alpha_1 \pi}{2}\cos\frac{\alpha_2 \pi}{2}, \;\;\; 
 a_2= \cos\frac{\alpha_2 \pi}{2}\sin \frac{\alpha_1 \pi}{2} e^{2 i \pi \varphi_1} ,\;\;\;
 a_3=\sin \frac{\alpha_2 \pi}{2} e^{2 i \pi \varphi_2},
 \end{eqnarray}
 where
 \begin{eqnarray}\label{alpint}
0\le \alpha_i  \le 1, \;\;0\le \varphi_i\le 1,\;\;i=1,2,
 \end{eqnarray}
 and the parameters $\alpha_i$, $\varphi_i$, $i=1,2$, are referred to as the control parameters.

\subsection{Local state of receiver}
The state of the two-qubit receiver 
at some time instant $t$ can be obtained reducing the state of the whole chain over spins $1,\dots,N-2$. 
Written  
 in the basis
\begin{eqnarray}\label{BB}
|0\rangle, \;\; 
|N-1\rangle,\;\;|N\rangle,\;\;|N (N-1)\rangle,
\end{eqnarray}
the receiver's density matrix reads as follows:
\begin{eqnarray}\label{rhoB}
\rho^R\equiv {\mbox{Tr}}_{1,2,\dots,N-2} \rho=\left(
\begin{array}{cccc}
1-|f_{N-1}|^2-|f_{N}|^2 & 0&0&0\cr
0&|f_{N-1}|^2 & f_{N-1} f_{N}^*&0\cr
0&f_{N-1}^* f_{N}& |f_{N}|^2 &0\cr
0&0&0&0
\end{array}
\right)
\end{eqnarray}
(in basis (\ref{BB}), the vector $|N (N-1)\rangle$ means the state with $N$th and $(N-1)$th excited spins).
Here 
star means the complex conjugate value and $f_{N-1}$, $f_N$, $f_0$ are the  transition 
amplitudes,
\begin{eqnarray}
f_i&=&\langle i| e^{-i H t} |\Psi_0\rangle = R_{i} e^{2 \pi i \Phi_i},\;\;i=0,\dots,N,
\end{eqnarray}
where $R_i$ and $\Phi_i$ are the real parameters and $R_i$ are positive.
Remember the  natural constraint  
 \begin{eqnarray}
\label{constr}
|f_N|^2 +|f_N-1|^2 \le 1 \;\; \Rightarrow \;\; R^2\equiv R_N^2 +R_{N-1}^2 \le 1,
\end{eqnarray}
where the equality corresponds to the perfect two-qubit state transfer
because in this case  $f_i \equiv 0 $ ($i<N-1$).

Obviously, the probability amplitudes appearing in the receiver's state (\ref{rhoB}) are  linear functions 
of the  parameters $a_i$:
 \begin{eqnarray}\label{NN}
f_N(t)&=&\langle N| e^{-i H t} |\Psi_0\rangle = \sum_{j=1}^3 a_j \langle N| e^{-i H t} |j\rangle
=\sum_{j=1}^3 a_j p_{Nj}(t)\\\label{NNm1}
f_{N-1}(t)&=&\langle N-1| e^{-i H t} |\Psi_0\rangle = \sum_{j=1}^3 a_j \langle N-1| e^{-i H t} |j\rangle
=\sum_{j=1}^3 a_j p_{(N-1)j}(t),
\end{eqnarray}
where
\begin{eqnarray}\label{def_chi}
  p_{kj}(t)=\langle k| e^{-iH t}|j\rangle = r_{kj}(t) e^{2 \pi i \chi_{kj}(t)},\;\;k,j>0,
 \end{eqnarray} 
$r_{kj}$ are the positive amplitudes  and $2 \pi \chi_{kj}$ ($0\le \chi_{kj}\le 1 $)  
are the phases of $p_{kj}$. 
 The meaning of $p_{kj}$ is evident. It is the transition amplitude of the excitation 
 from the $j$th to the $k$th spin. 
 Emphasize that these amplitudes  represent the 
 inherent characteristics of the transmission
 line and do not depend on the control parameters of the sender's initial state.

 \section{Quantum correlations at  receiver side}
 \label{Section:Q}
 We introduce two parameters characterizing the quantum correlations at the receiver side.
The first of these parameters is 
 the discord $Q_{ext}$ between the 
 receiver and the rest of a chain, it indicates whether these two subsystems correlate  to one  another. The second one 
is the discord  $Q_R$ between the qubits of the receiver, it characterizes the  correlations inside of the receiver.

 \subsection{Discord between the receiver and the rest of communication line}
 Since the initial state of our system is a pure one, it remains pure during the evolution.  
Thus,  the receiver and the rest of communication line compose the whole system in a pure state. 
Consequently, the discord between these two subsystem is identical to the entanglement between them \cite{DSC}. 
The later can be simply calculated in terms of the entropy:
\begin{eqnarray}
Q_{ext}=-{\mbox{Tr}} \rho^R \log_2 \rho^R = -\sum_{i=1}^4 \lambda_i \log_2\lambda_i.
\end{eqnarray}
In our case, $\rho^R$ (\ref{rhoB}) is an  $X$-matrix 
having the following two nonzero eigenvalues:
\begin{eqnarray}
\lambda_1=1-R^2,\;\;\lambda_2=R^2.
\end{eqnarray}
Consequently, 
\begin{eqnarray}\label{Qext}
Q_{ext}=-\sum_{i=1}^2 \lambda_i \log_2\lambda_i = - R^2\log R^2 - (1-R^2)\log(1-R^2).
\end{eqnarray}

\subsection{Inside discord of the receiver}
The formula for 
discord between the two nodes of receiver is more complicated. For the particular case of  $X$-matrix (\ref{rhoB}) 
it was derived in  \cite{FZ_QIP2014}  (see 
Appendix for more details):
\begin{eqnarray}\label{final_discord_ex}
Q_{R}=\min(Q_{N},Q_{N-1}),
\end{eqnarray}
where
\begin{eqnarray}\label{final_discord_ex0}
Q_{N}&=&1 -R_{N-1}^2 \log_2 R_{N-1}^2 -\\\nonumber
&& (1-R_{N-1}^2)\log_2(1-R_{N-1}^2)
+R^2 \log_2 R^2+\\\nonumber
&&
 (1-R^2)\log_2(1-R^2)  -\\\nonumber
 &&\frac{1}{2}
 \Big(1-\sqrt{1-4R_{N}^2(1-R^2)}\Big)\log_2
(1-\sqrt{1-4R_{N}^2(1-R^2)}) -
 \\\nonumber
 &&
 \frac{1}{2}
 \Big(1+\sqrt{1-4R_{N}^2(1-R^2)}\Big)\log_2
(1+\sqrt{1-4R_{N}^2(1-R^2)}),\\\nonumber
Q_{N-1}&=&Q_N|_{N-1 \leftrightarrow N}.
\end{eqnarray}
This discord   depends on the 
absolute values $R_{N}$, $R_{N+1}$  of the transition amplitudes:

\subsection{$R$- and $R_{N-1}$-dependence of  discords $Q_{ext}$ and $Q_R$} 
\label{Section:analysQ}
\begin{figure*}
   \epsfig{file=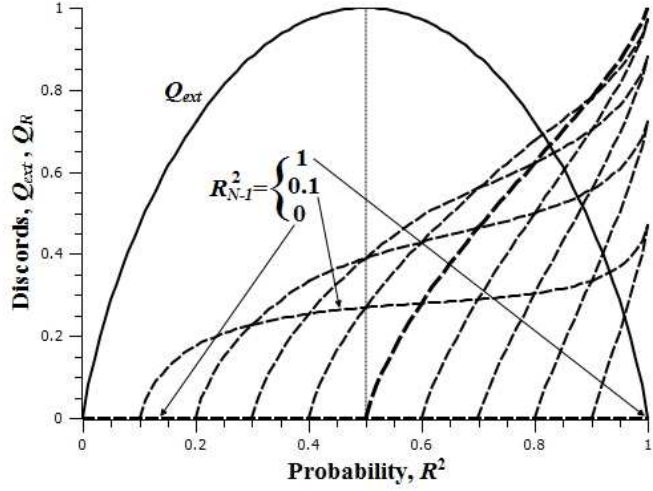,
  scale=0.5
   ,angle=0
}
\caption{The discords $Q_{ext}$ (solid line) and $Q_R$ (dash-line) as functions of $R^2$ and $R_{N-1}^2$. 
The different dash-lines correspond to the different values of $R_{N-1}^2 = 0.1 n$, $n=0,1,\dots,10$. The dash-line 
$R^2_{N-1}=0$ coincides with the abscissa axis, the line $R^2_{N-1}=1$ shrinks to the point $R^2=1$.
The bold dash-line corresponds to 
$R_{N-1}^2=\frac{1}{2}$,   the discord  $Q_R$ can take any allowed value on this line, $0\le Q_R\le 1$.
} 
  \label{Fig:QQ} 
\end{figure*}

In Fig.\ref{Fig:QQ}, we represent the discord $Q_{ext}$ as a function of $R^2$  (the solid bell-shaped line)
and $Q_{R}$ as a function of $R^2$ and $R_{N-1}^2$ (the family of dash-lines, each line corresponds to the particular 
value of $R_{N-1}^2$). 
 In this graph we see that the large values of the discord $Q_R$ can be produced by  large $R$. Therewith,
  the maximal value of the discord $Q_R$, 
 ($Q^{max}_{R}=1$) corresponds to $R^2=1$ and
 $R_{N-1}^2=\frac{1}{2}$ (the bold dash-line). 
 On the contrary, the
 discord $Q_{ext}$ has the maximum at $R^2=\frac{1}{2}$, this means 
 that $Q_R$ decreases with either $R_{N-1}\to 0$ or $R_{N-1}\to 1$. 
 Remark, that  there is a region in this figure where $Q_{R}$ is 
 large while $Q_{ext}$ is rather small
 (the right upper corner of the figure). 
In this region the quantum correlations between the  receiver and the rest of communication line 
are minimized and therefore  the receiver can be used as (almost) independent object. 
However, this region is difficult for realization and can be created when the 
chain is engineered for the high probability state transfer.

\section{Time optimization of remote  quantum correlations}
\label{Section:opt}
The remote control of quantum states is aimed at creation of required parameters at the receiver side by varying the
control parameters. 
Formally, there is analytical relation between the control parameters and creatable ones. Moreover, the elements of the 
receiver's density matrix are linear functions of the  parameters $a_i$ as was mentioned above. However,
the coefficients of these linear functions  depend on the transition amplitudes $p_{kj}$ (\ref{def_chi})
(the $t$-dependent inherent characteristics of the transmission line) and thus are hardly understandable
without   graphic representation. Therefore  below we  numerically study  the  map  
of the domain of the  two control parameters $\alpha_1$ and $\alpha_2$ into the plane of the 
creatable parameters $Q_{ext}$ and $Q_R$:
\begin{eqnarray}
\label{map}
(\alpha_1,\alpha_2) \to (Q_{ext},Q_R).
\end{eqnarray}
Note that we set $\varphi_i=0$ in formulas (\ref{aalpha}) for $a_i$ because the effect of these 
phases  is negligible in our model, this conclusion was confirmed by the preliminary numerical simulations.

Using the parameter $\phi$ in eq.(\ref{Dnonhom}) we vary the chain from the ideal Ekert chain ($\phi=\frac{1}{2}$, 
the whole receiver's 
state-space can be created in terms of $Q_{ext}$ and $Q_R$ in this case)  
to the homogeneous one ($\phi=0$, the creatable region  is  minimal in this case).

\subsection{Time optimization  of discords $Q_{ext}$ and $Q_R$}

Taking into account  formulas (\ref{Qext}) and (\ref{final_discord_ex}) and the discussion in 
Sec.\ref{Section:analysQ} we conclude that 
the probability of the state transfer to the receiver side, $R^2$, is the most relevant parameter responsible for the 
quantum correlations and must be studied 
in more detail.

According to formula   (\ref{Qext}) the discord $Q_{ext}$ 
vanishes as either $R=0$ or $R=1$. 
In the ideal  case, $R=1$,  the  signal is completely collected at the nodes of  the receiver. 
However, usually $R<1$ and depends on the initial state of the spin system. 
In the next subsection we perform the time-optimization of $R$ for the 
 initial state (\ref{sins},\ref{aalpha}) with the particular values of control parameters:
$\alpha_i=0$, $i=1,2$.

\subsubsection{Time optimization  of state transfer probability $R^2$}
\label{Section:R}
 
\begin{figure*}
   \epsfig{file=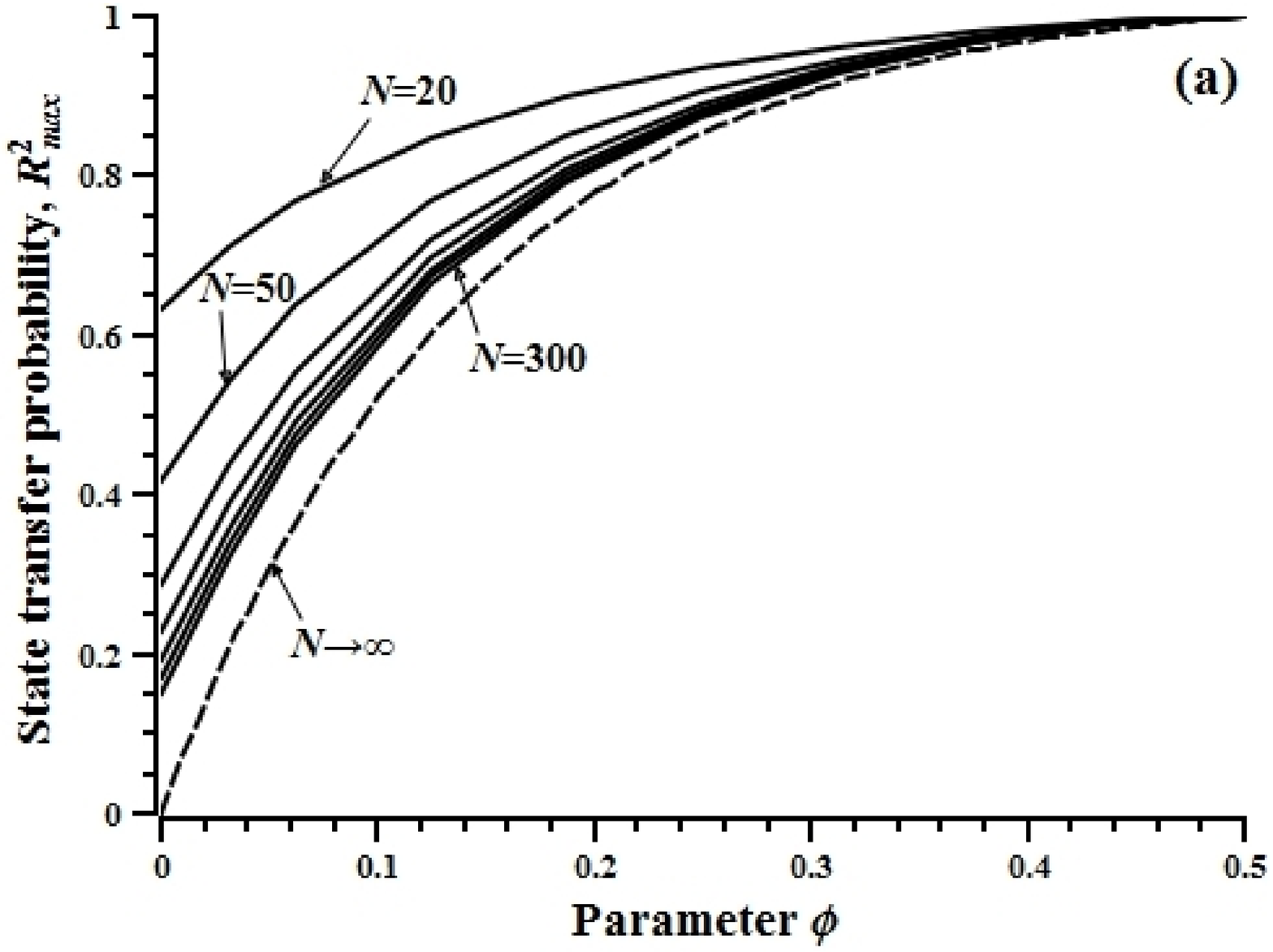,
  scale=0.45
   ,angle=0
}\epsfig{file=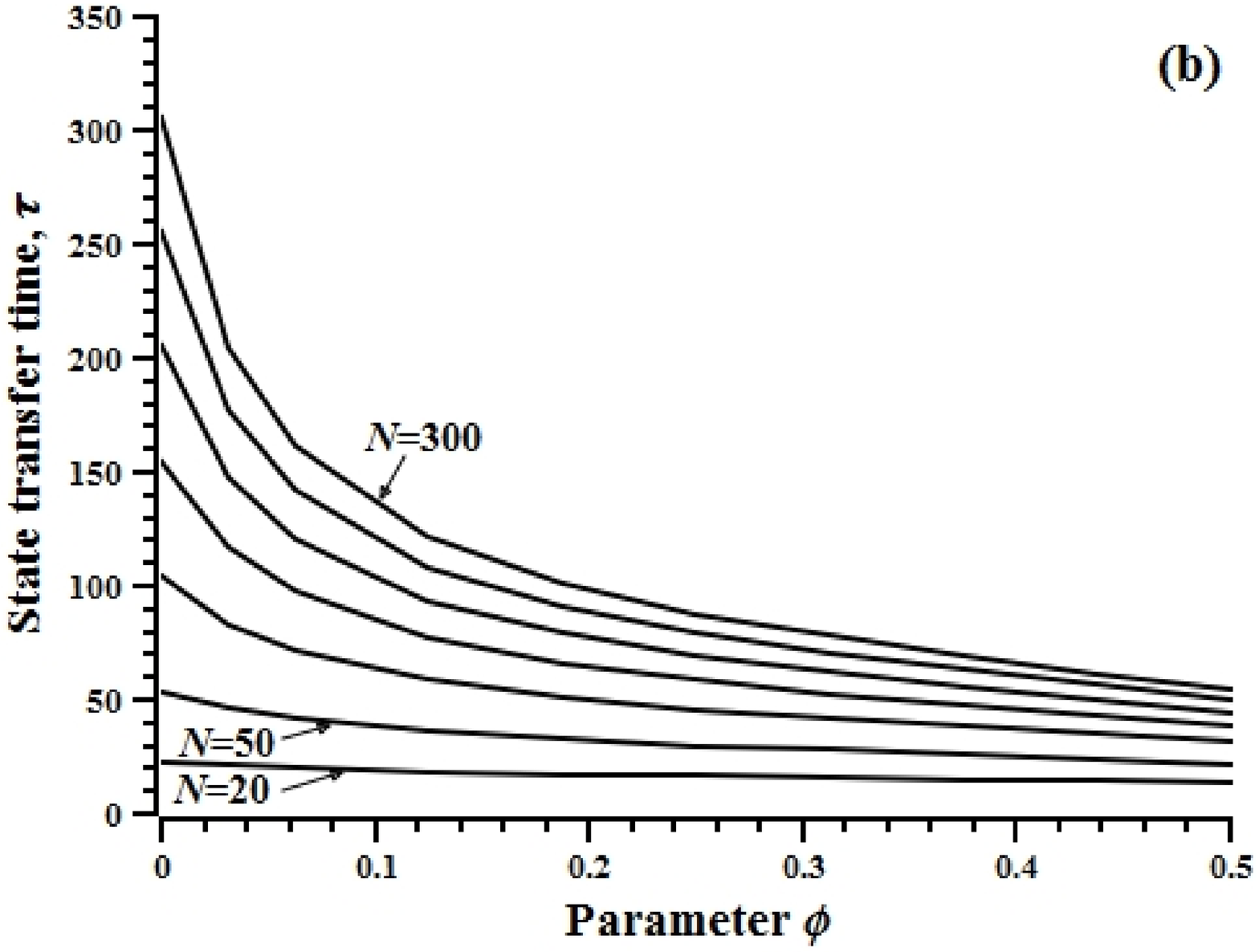,
  scale=0.45
   ,angle=0
}

\caption{The maximum of the state transfer probability $R^2$ and the appropriate time instant $\tau$ for the 
initial state  $\Psi_0=|1\rangle$ and chains of different lengths, $N=20, 50 n$, $n=1,2,\dots,6$. 
} 
  \label{Fig:destr} 
\end{figure*}

The probability $R^2$ as a function of the time $t$ is 
an oscillating function of time and reaches the 
first maximum  $R_{max}^2$  (the largest one) at some time instant $t_{0}$. 
Both of these parameters ($R_{max}^2$ and  $t_{0}$) are shown in Fig.\ref{Fig:destr}
as functions of the inhomogeneity parameter $\phi$ for the chains of different lengths  $N$,
\begin{eqnarray}
N=20,\;50 n,\;\;n=1,\dots,6.
\end{eqnarray}
Fig.\ref{Fig:destr}a  shows that the amplitude approaches unit as 
$\phi \to \frac{1}{2}$ 
(Ekert chain). 
There is a  limiting curve $N\to\infty$  in Fig.\ref{Fig:destr}b (dash-line) 
showing that the state creation 
algorithm  becomes more $N$-independent
with approaching to the Ekert case, $\phi\to\frac{1}{2}$, because all curves approach each other 
in the right upper corner
of this figure. 

To obtain the approximate form of the limiting curve in Fig.\ref{Fig:destr}, we note that 
each curve  in Fig.\ref{Fig:destr}a can be approximated by the function
\begin{eqnarray}
F_N=c_N - \exp( - a_N \phi \pi - b_N ),
\end{eqnarray}
with particular values of the coefficients $a_N$, $b_N$ and $c_N$ (we do not represent these curves 
in Fig.\ref{Fig:destr}a, we also do not give the values of the parameters $a_N$, $b_N$ and $c_N$ for  brevity).
Studying the dependence of the parameters  $a_N$, $b_N$ and $c_N$ on $N$  
we observe that $a_N$ has the  well-formed  asymptotics as 
$N\to\infty$: $a_\infty \approx 2.232$.
The two other parameters $b_\infty$ and $c_\infty$ can be approximated  using the ''boundary'' requirements $
F_\infty|_{\phi=\frac{1}{2}}=1$ and $F_\infty|_{\phi=0}=0$: $b_\infty \approx  -0.03$, $c_\infty\approx  1.031$.
Thus, we approximate the limiting curve (the dash-line in Fig.\ref{Fig:destr}a) by the function 
\begin{eqnarray}
R_\infty=1.031 - e^{-2.232 \phi \pi + 0.03}.
\end{eqnarray}
As for  the time instant $t_0$, it  increases linearly with $N^{\gamma(\phi)}$ ($t \sim N^{\gamma(\phi)}$), where   $\gamma$ 
decreases  with increase in $\phi$ from $\gamma(0)=1$ to $\gamma(\frac{1}{2})=\frac{1}{2}$.

 \section{Numerical simulations of map (\ref{map})}
 \label{Section:num}
 The main purpose of  numerical simulation  of  map (\ref{map}) is 
  revealing  the dependence of the area of  creatable region 
 on the inhomogeneity parameter $\phi$. In particular, we select  the sub-domain 
 in the control-parameter space for which  map
 (\ref{map}) is (almost) the  one-to-one  map.
 
 \subsection{Domain of control parameters}
 For convenience, we separate the whole domain of  control parameters (\ref{alpint}) into the four sub-domains 
 (we put $\varphi_i=0$ for the reason indicated above). 
 
\noindent 
The first sub-domain:
\begin{eqnarray}\label{reg}
 0\le\alpha_i\le \frac{1}{2},\;\;\;i=1,2. 
 \end{eqnarray}
 The second sub-domain:
\begin{eqnarray}\label{reg12}
\frac{1}{2} \le\alpha_1\le 1,\;\;
0\le\alpha_2\le \frac{1}{2}.
\end{eqnarray}
The third sub-domain:
\begin{eqnarray}\label{reg21}
0 \le\alpha_1\le \frac{1}{2}, \;\;\; \frac{1}{2} \le\alpha_2\le 1.
\end{eqnarray}
The fourth sub-domain:
\begin{eqnarray}\label{reg22}
\frac{1}{2} \le\alpha_i\le 1,\;\;\;i=1,2.
\end{eqnarray}
The reasoning for this separation is clarified below in Secs.\ref{Section:Ekert}, \ref{Section:nonhom}.
 It will be shown that sub-domain (\ref{reg}) is (almost) one-to-one mapped into the creatable region.

\subsection{Ekert chain} 
\label{Section:Ekert}
\begin{figure*}
\epsfig{file=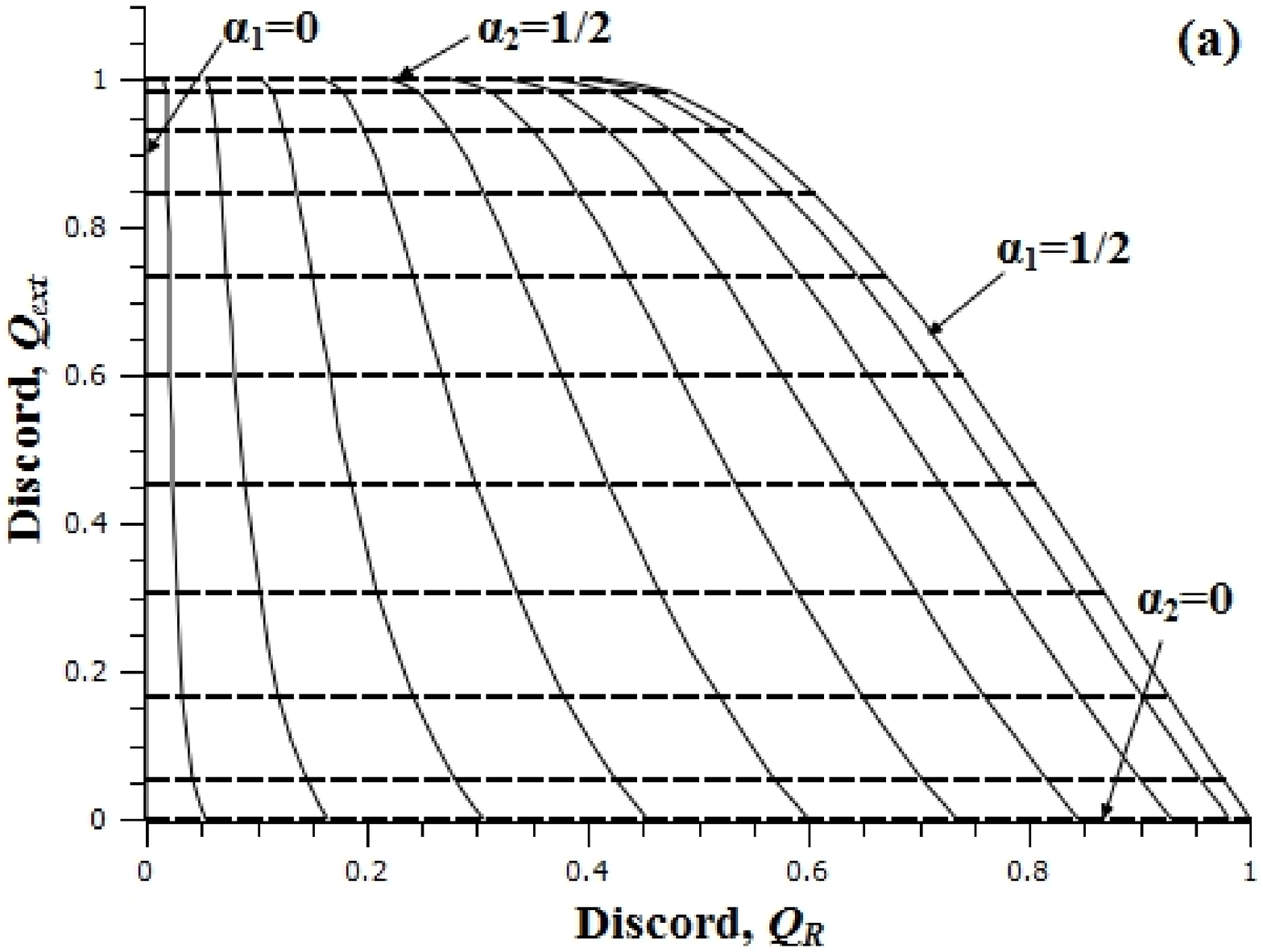,
  scale=0.45
   ,angle=0
}
\epsfig{file=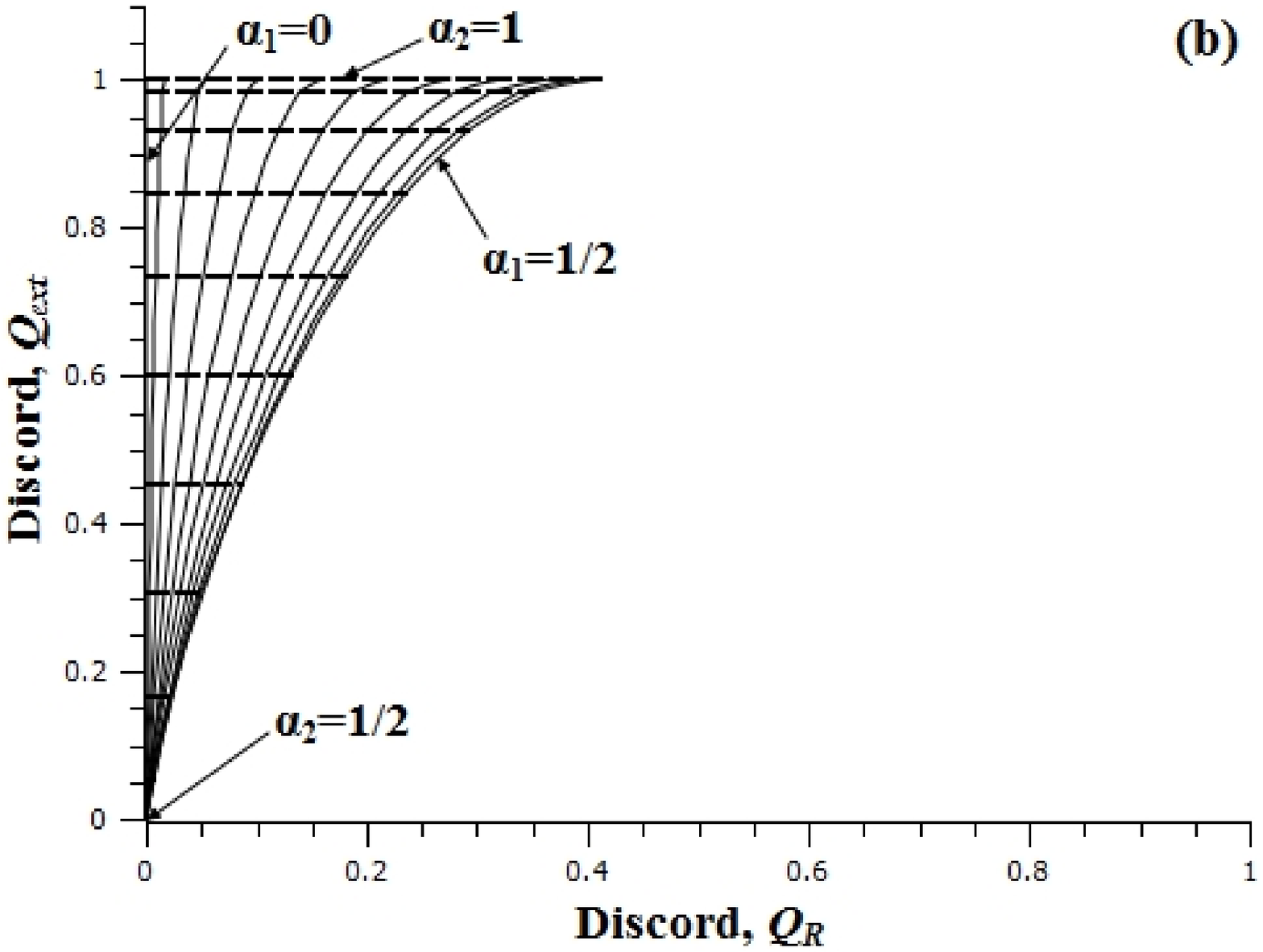,
  scale=0.45
   ,angle=0
}
\caption{The discord $Q_{ext}$ verses the discord $Q_{R}$ 
for the chain of $N=20$ nodes with the homogeneity parameter
$\phi=\frac{1}{2}$
(Ekert chain) at the time instant $t_0=13.69$. The vertical solid lines  and the  horizon dash-lines correspond to 
$\alpha_1=const$ and $\alpha_1=const$ respectively. The distance between the neighboring lines of each family 
is 0.05.
(a) The control parameters   $\alpha_i$, $i=1,2$, from  sub-domain (\ref{reg}). 
Sub-domain of  control parameters (\ref{reg12}) 
maps into the same region with $\alpha_1$ increasing from the right to the left reaching 
$\alpha_1=1$ at the left gridding line (the ordinate axis).
(b) The control parameters   $\alpha_i$, $i=1,2$, from  sub-domain  (\ref{reg21})
Sub-domain (\ref{reg22}) 
cover the same part of the creatable region with $\alpha_1$ 
increasing from the right to the left reaching  $\alpha_1=1$ at the left gridding line (the ordinate axis).
} 
  \label{Fig:Ekert} 
\end{figure*}

 In the limit case of the fully engineered Ekert  chain ($\phi=\frac{1}{2}$ in eq.(\ref{Dnonhom}))
  we are able to cover the whole space of 
 the parameters $Q_{R}$, $Q_{ext}$, see Fig.\ref{Fig:Ekert} where $N=20$. In this figure, the horizontal dash-lines correspond to 
 $\alpha_2=const$, while the solid lines correspond to  $\alpha_1=const$. 
Emphasize that it is not necessary to work with the  whole domain (\ref{alpint}) of  control parameters because the 
parameters from the  first sub-domain (\ref{reg})  cover the whole creatable space, as is shown in Fig.\ref{Fig:Ekert}a,
where 
some  particular values of the control parameters are indicated. Therewith the map (\ref{map}) 
is one-to-one map for this sub-domain. The parameters from the second sub-domain (\ref{reg12})
are mapped into  the same  region in Fig.\ref{Fig:Ekert}a (therewith, the parameter $\alpha_1$ increases
from  $\frac{1}{2}$ to  1 in passing from the  right to the left.
We shall point on the third (\ref{reg21}) and the fourth (\ref{reg22}) sub-domains. Both of them are mapped 
into the creatable subregion 
shown in Fig.\ref{Fig:Ekert}b. The indicated values of the control parameters $\alpha_i$ correspond to the third sub-domain 
(\ref{reg21}). The sub-domain (\ref{reg22}) maps into the same sub-region with $\alpha_1$  increasing from  $\frac{1}{2}$ 
to  1 in passing  from the right to the left.

 Thus, the subregion in Fig.\ref{Fig:Ekert}b is covered four times by the parameters from the
 all four sub-domains (\ref{reg}-\ref{reg22}) and consequently the states from this sub-region are simpler 
 creatable than others. This subregion correspond to the relatively small values of $Q_R$.  
 
All possible relations between the parameters $Q_{ext}$ and $Q_R$ are realizable in the Ekert case. In particular, 
the right upper corner 
in Fig.\ref{Fig:QQ} is mapped into the right lower corner in Fig.\ref{Fig:Ekert}.

 \subsection{Chains with $\phi<\frac{1}{2}$}
\label{Section:nonhom}
\begin{figure*}
 \epsfig{file=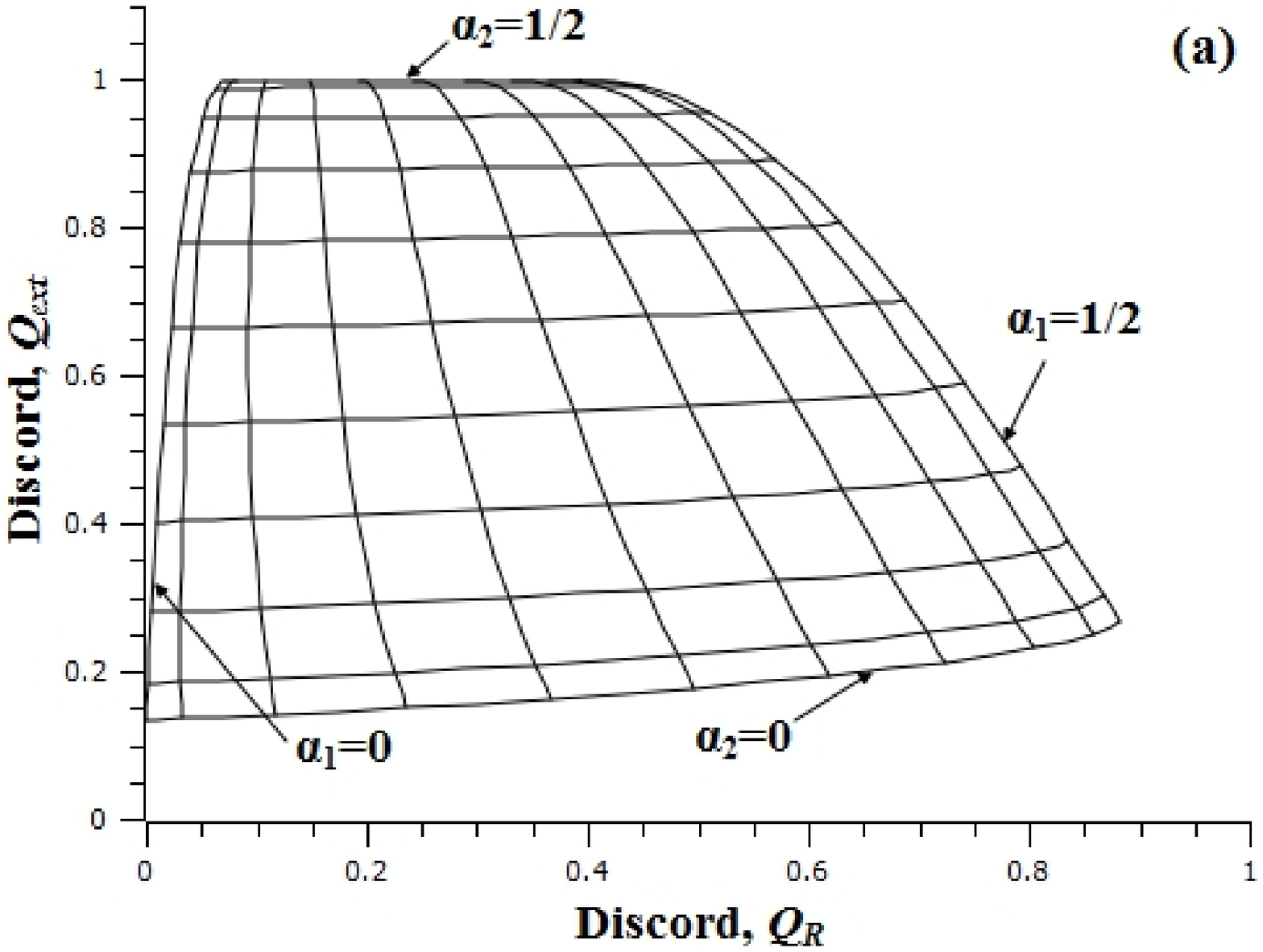,
  scale=0.35
   ,angle=0
}
\epsfig{file=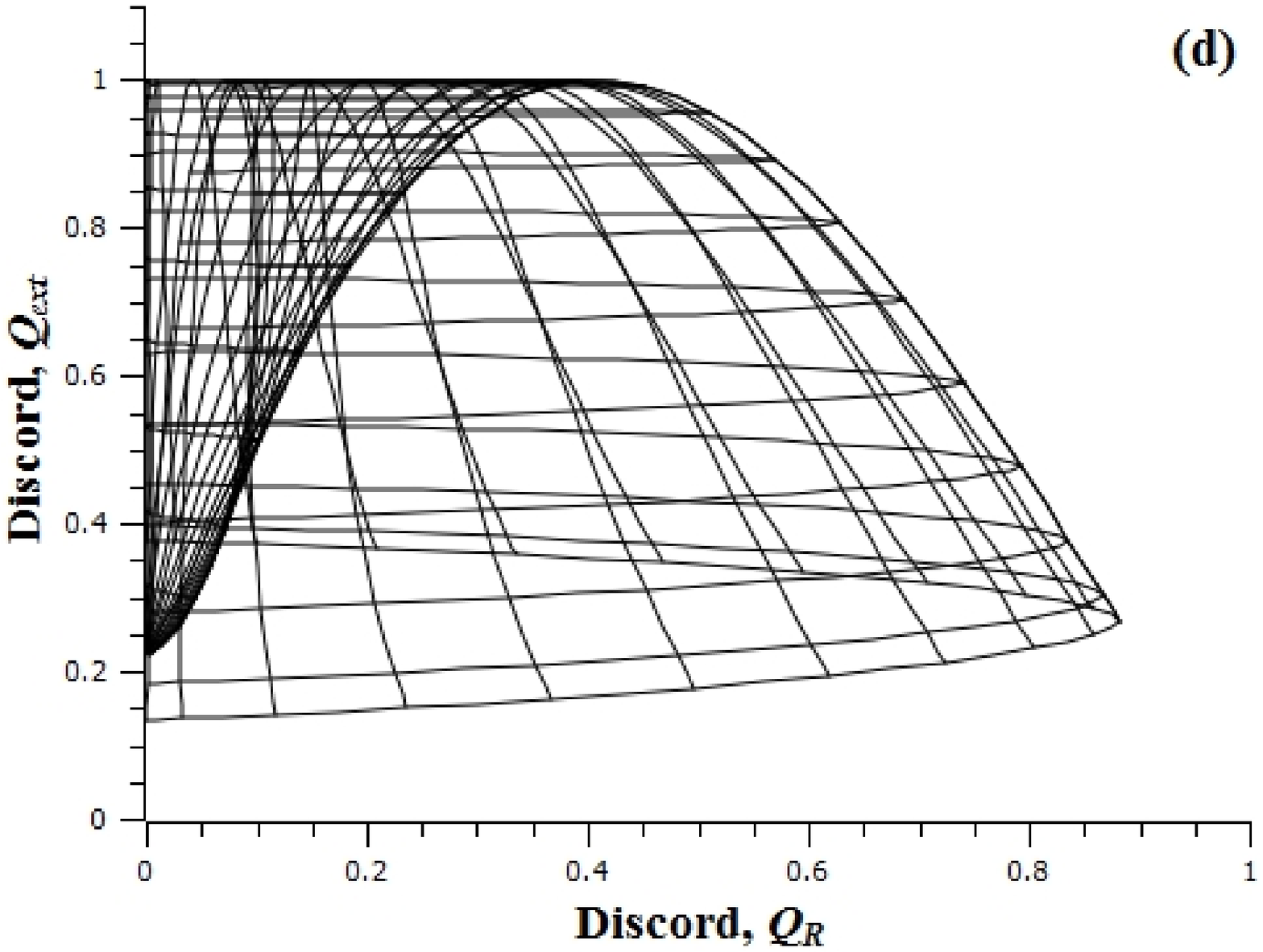,
  scale=0.35
   ,angle=0
}
\epsfig{file=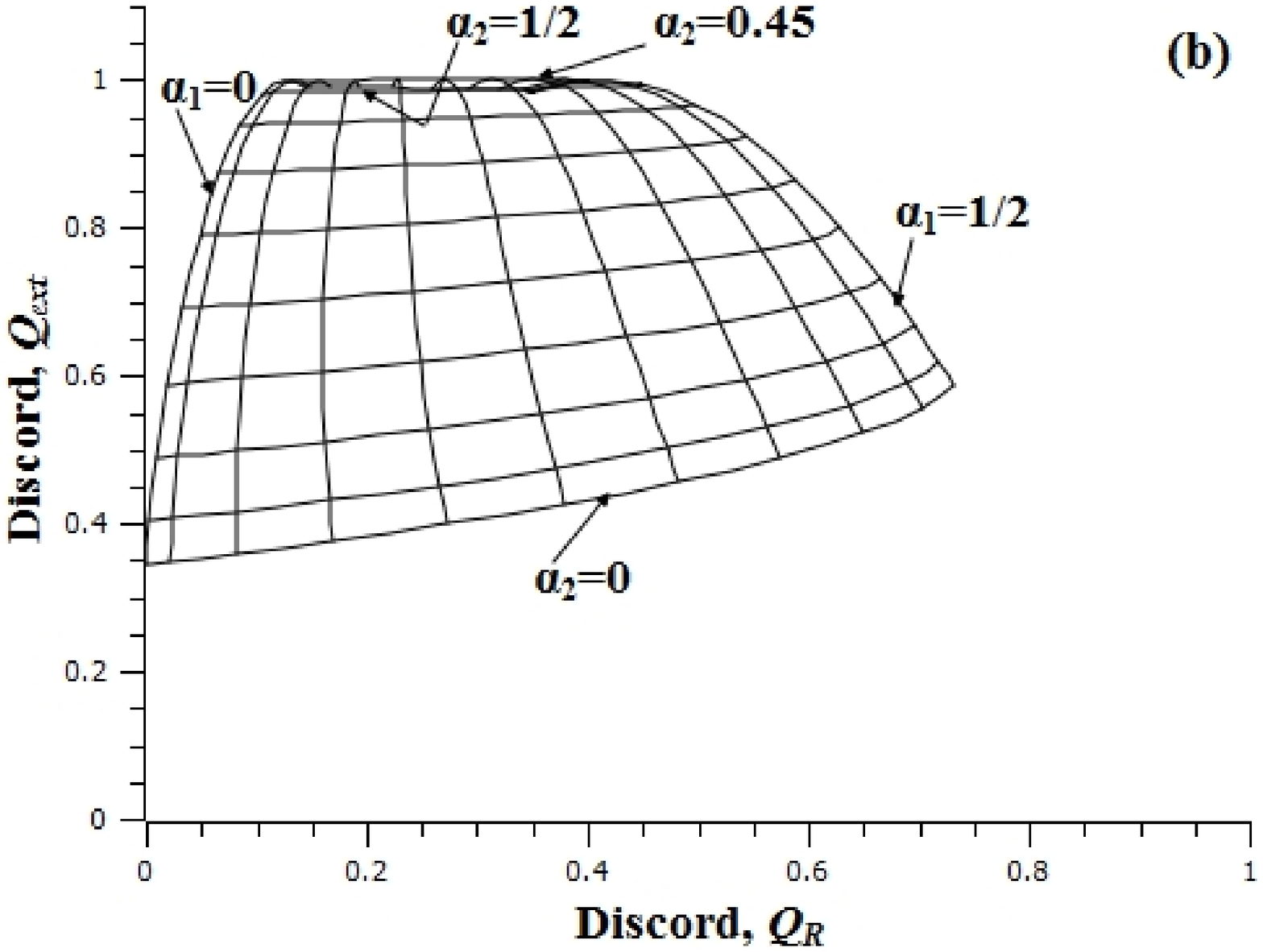 ,
  scale=0.35
   ,angle=0
}
\epsfig{file=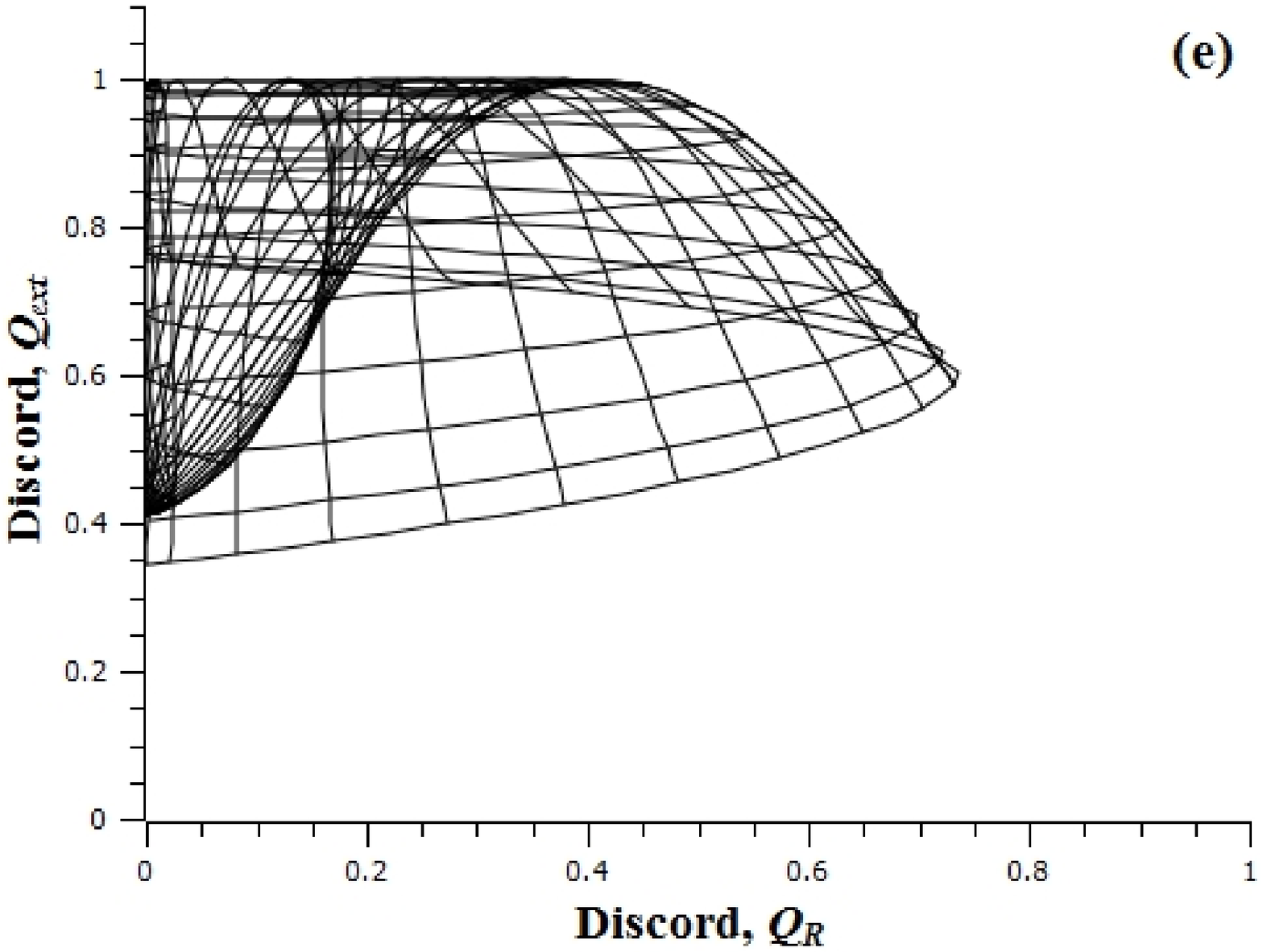,
  scale=0.35
   ,angle=0
}
 
   \epsfig{file=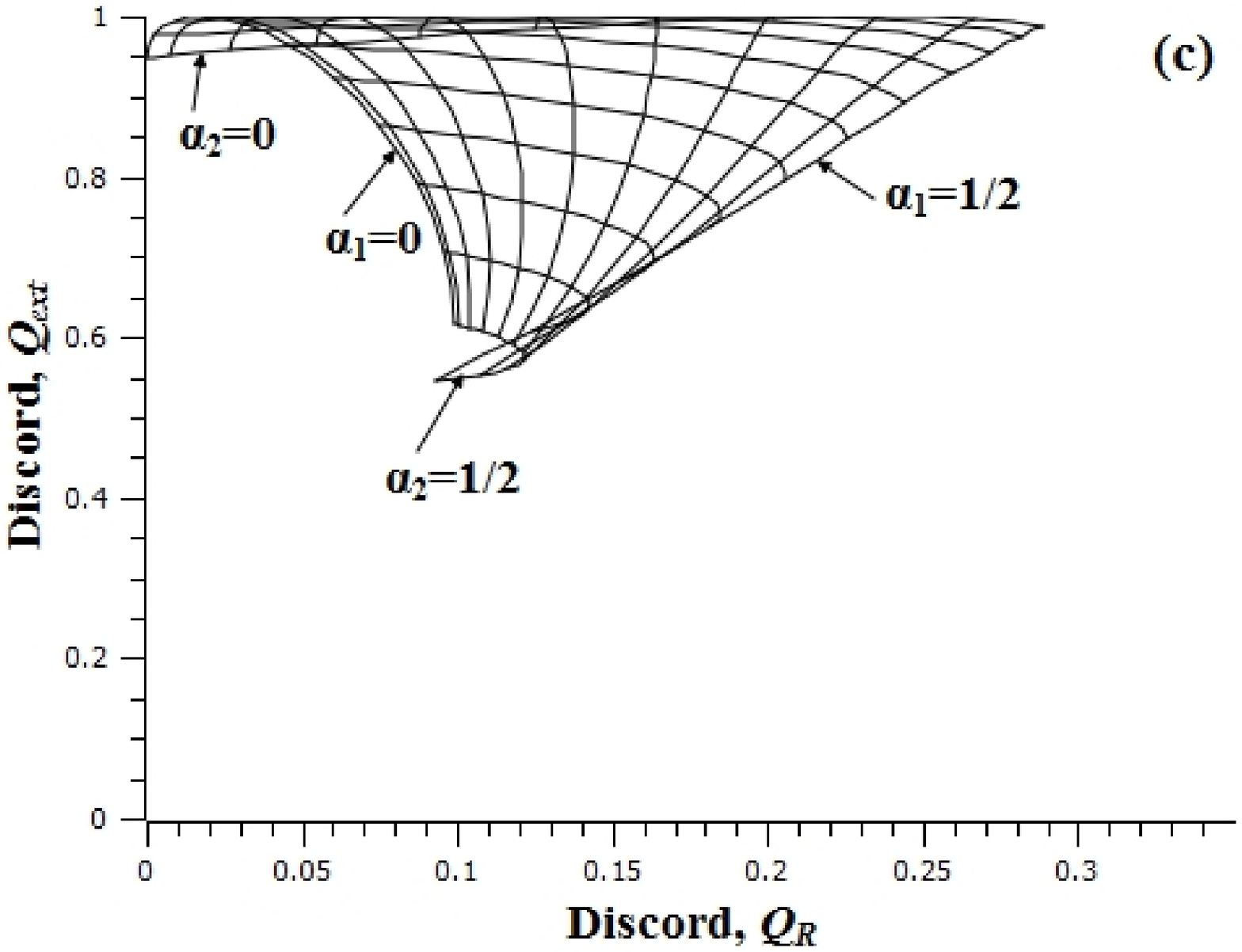,
  scale=0.35
   ,angle=0
}
\epsfig{file=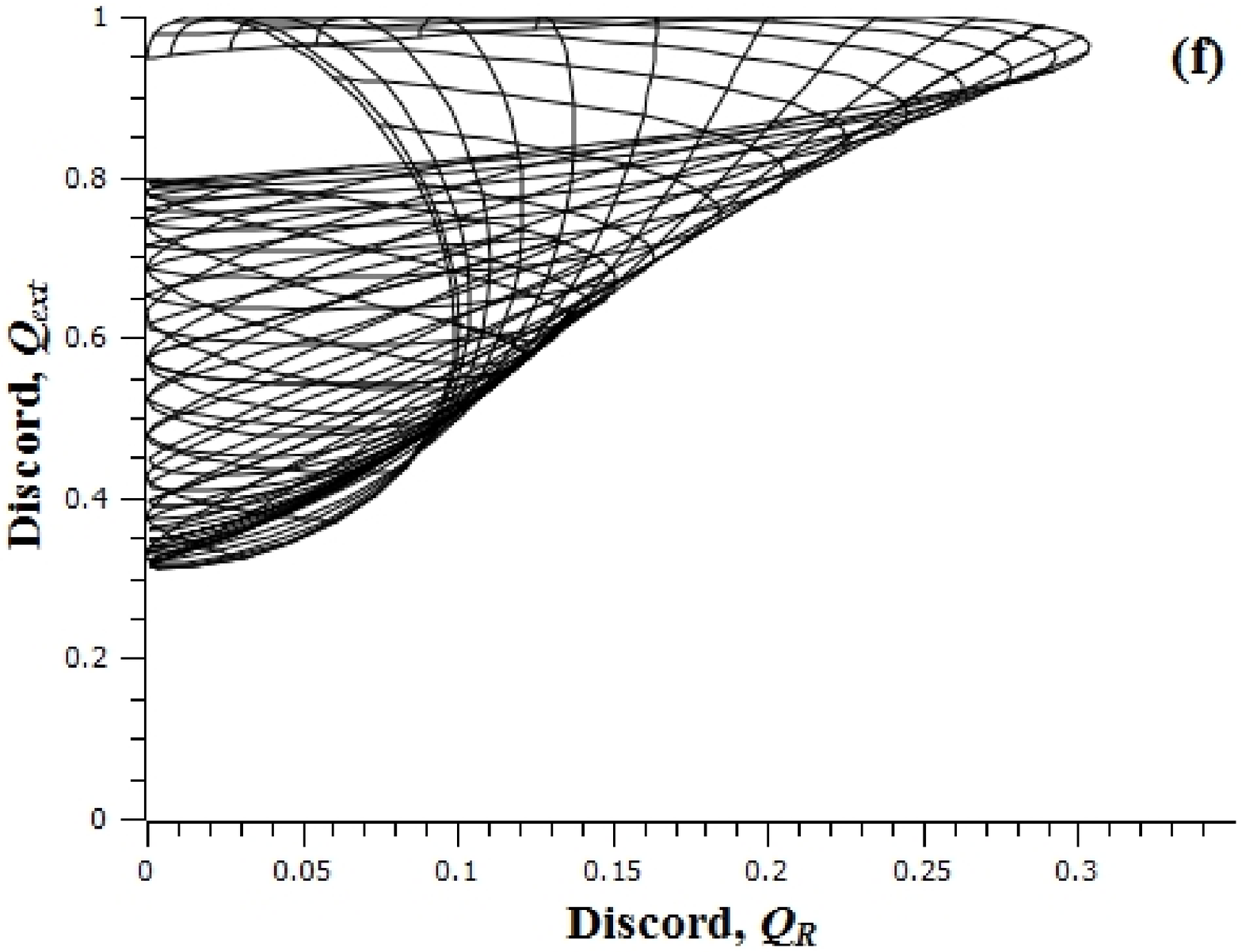,
  scale=0.35
   ,angle=0
}

\caption{The discord $Q_{ext}$ verses the discord $Q_{R}$ for the chain of 
$N=20$ nodes. The two crossing families of lines 
correspond to $\alpha_1=const$ and $\alpha_2=const$, similar to Fig.\ref{Fig:Ekert}.
The interval between the neighboring lines is $0.05$ (dimensionless units).
(a,b,c) The parameters  $\alpha_i$, $i=1,2$ vary inside of
sub-domain (\ref{reg}) of parameters  $\alpha_i$, $i=1,2$; (d,e,f) 
The parameters $\alpha_i$, $i=1,2$, vary 
inside of the whole domain (\ref{alpint}). (a,d) $\phi=\frac{3}{8}$, $t_0=15.27$, 
$R^2_{max}=0.98$; 
(b,e) $\phi=\frac{1}{4}$, $t_0=16.75$, $R^2_{max}=0.94$;
(c,f) $\phi=0$ (homogeneous chain), $t_0=22.79$, $R^2_{max}=0.63$.
} 
  \label{Fig:Num20} 
\end{figure*}

\begin{figure*}
\epsfig{file=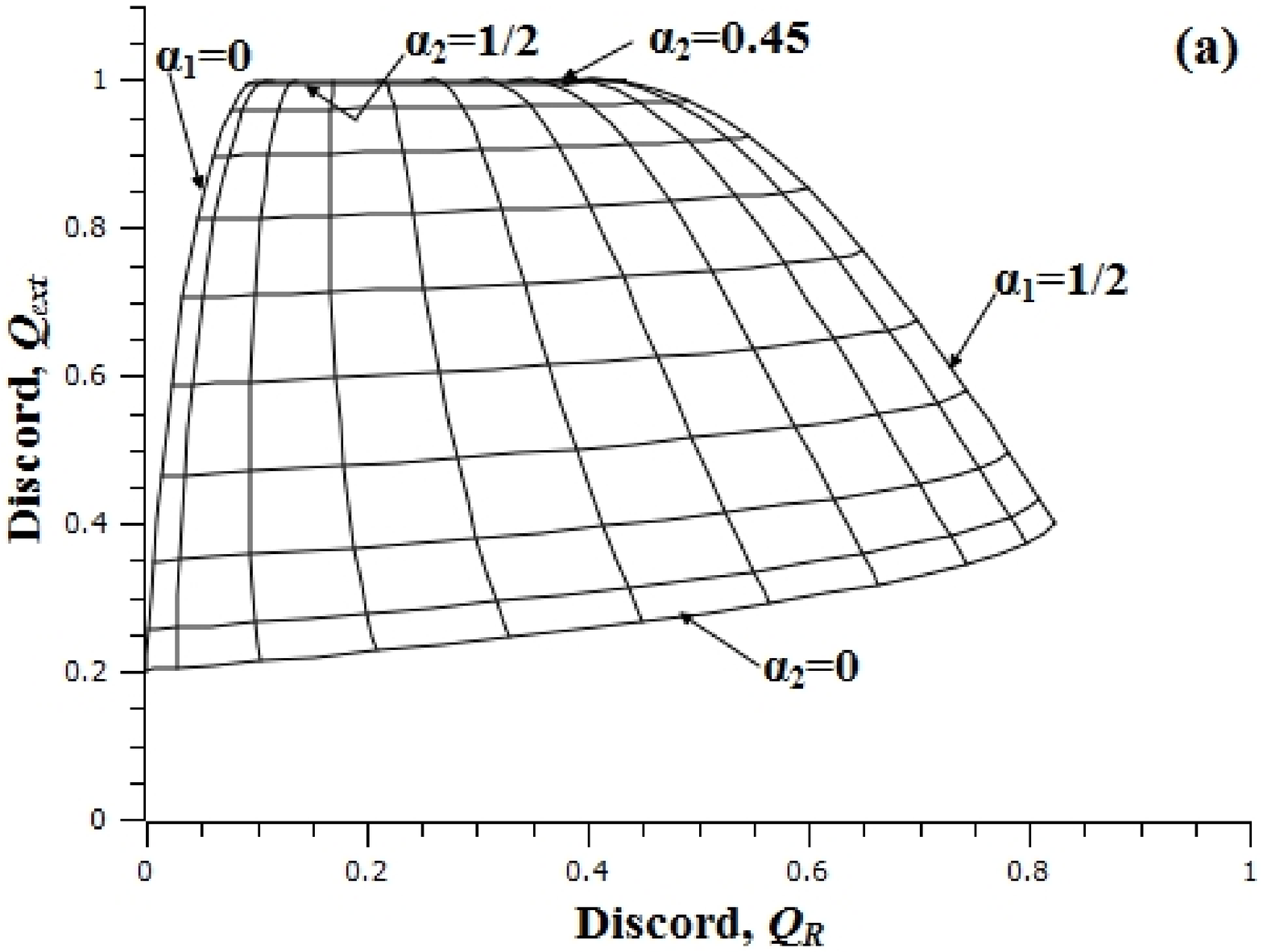 ,
  scale=0.35
   ,angle=0
}
\epsfig{file=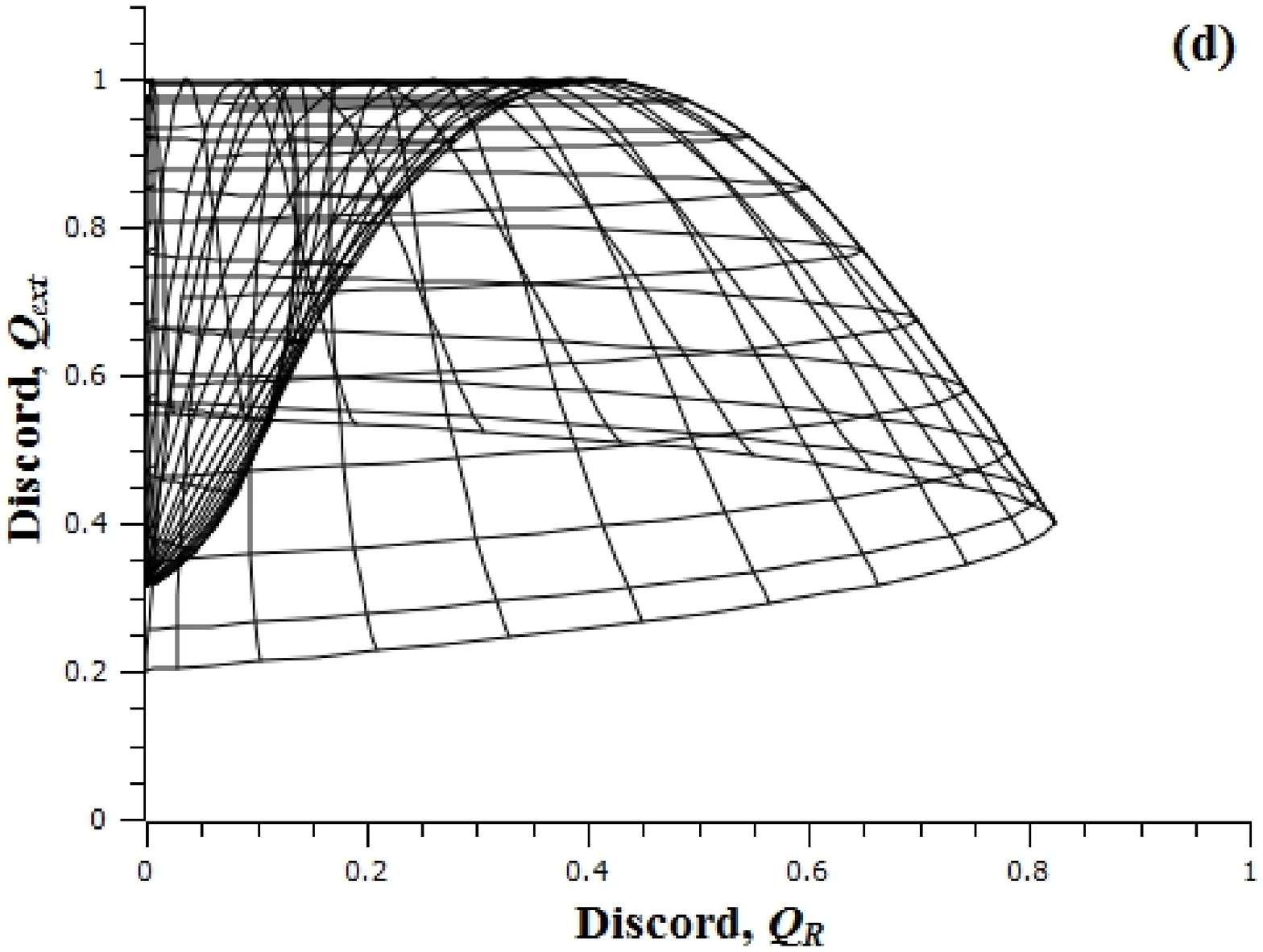,
  scale=0.35
   ,angle=0
}

\epsfig{file=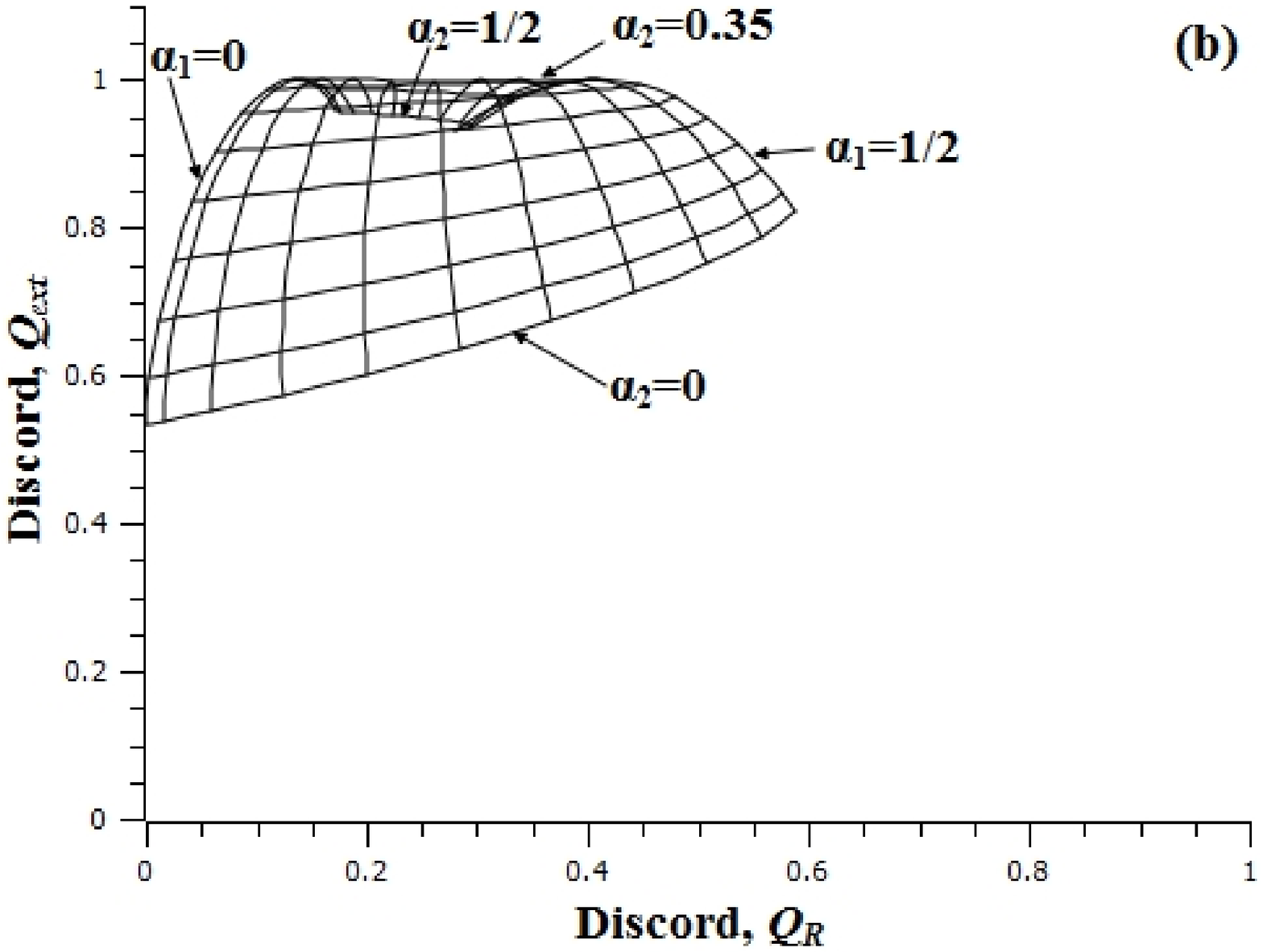,
  scale=0.35
   ,angle=0
}
\epsfig{file=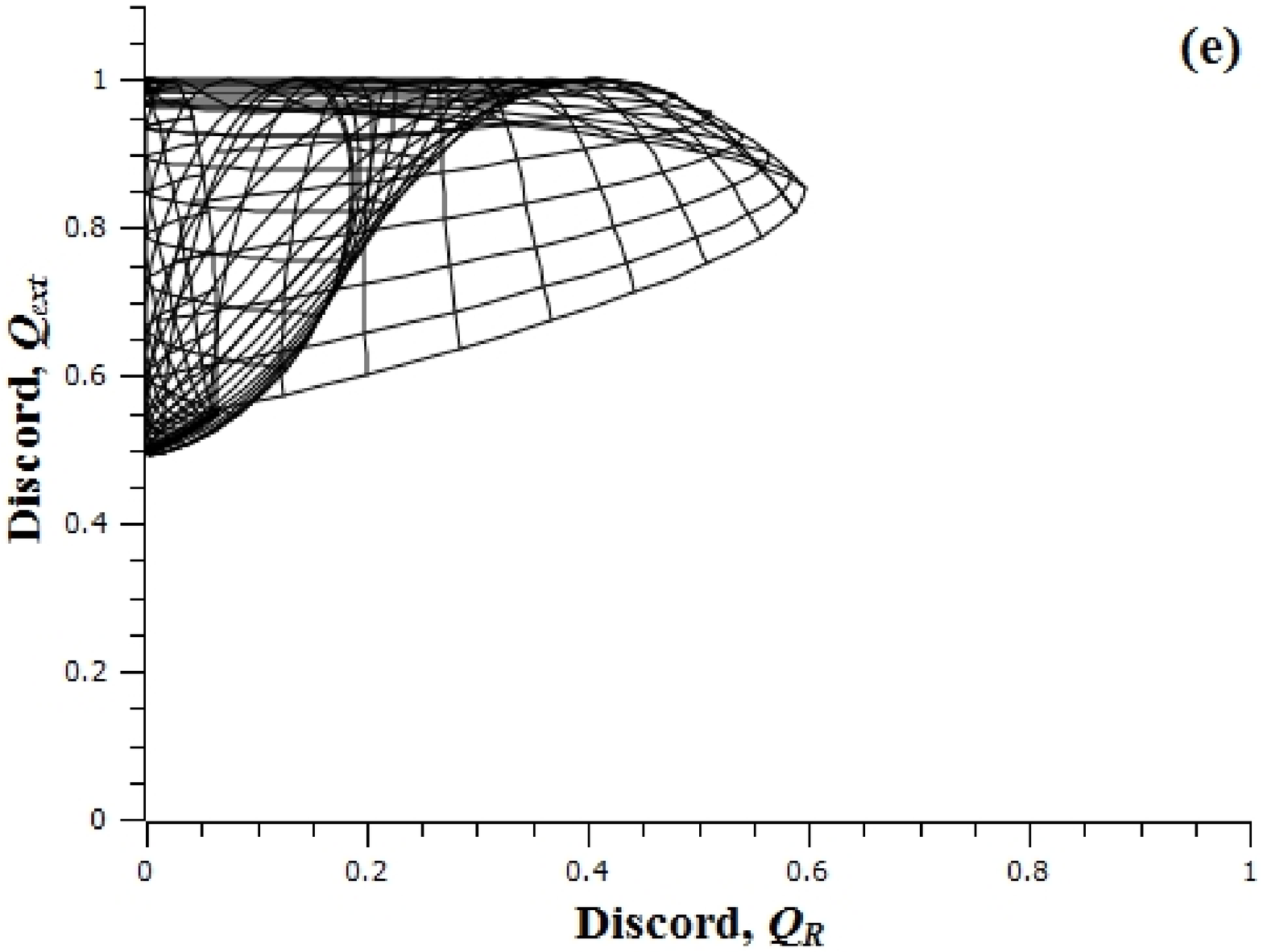,
  scale=0.35
   ,angle=0
}
   \epsfig{file=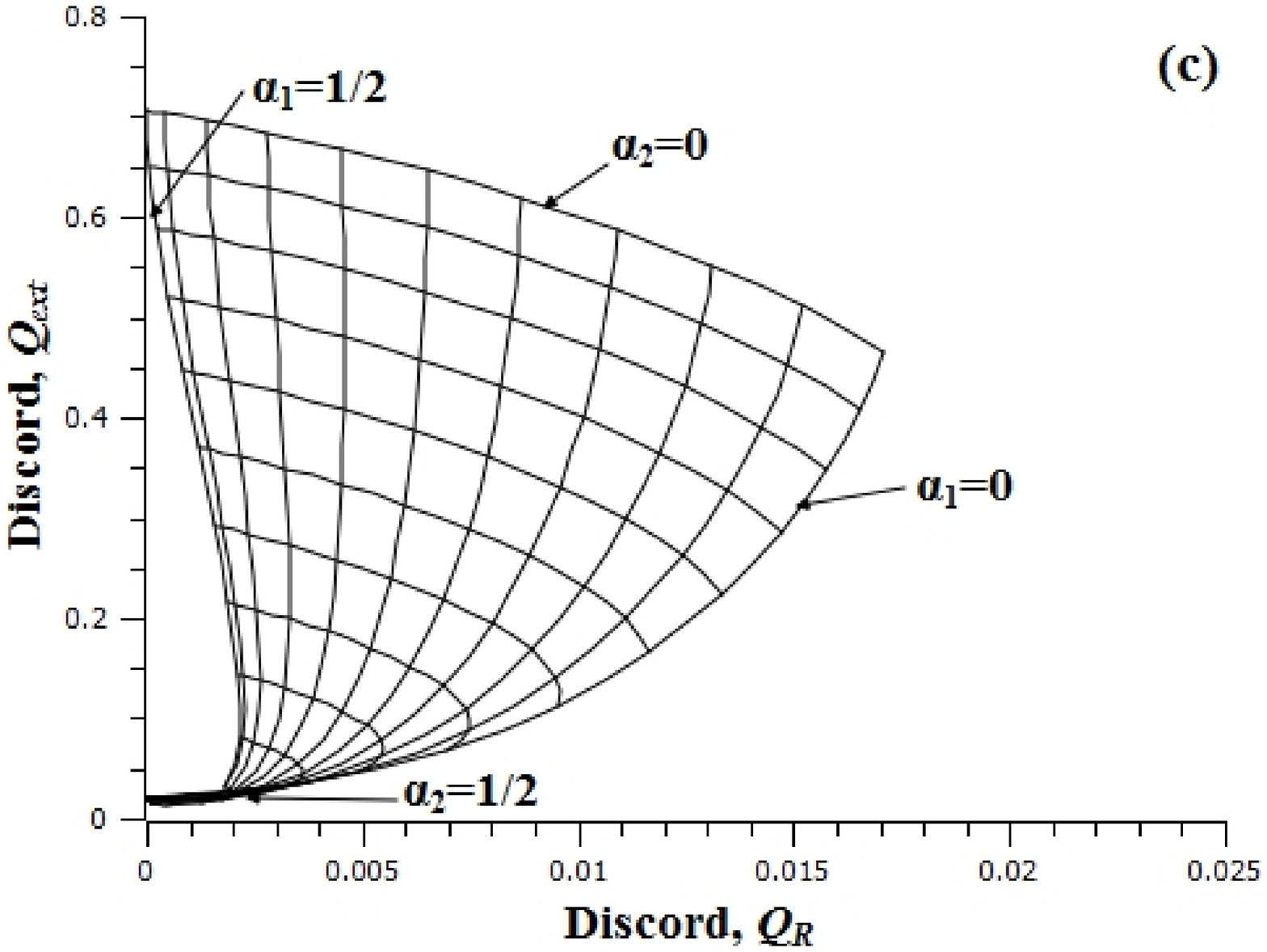,
  scale=0.35
   ,angle=0
}
\epsfig{file=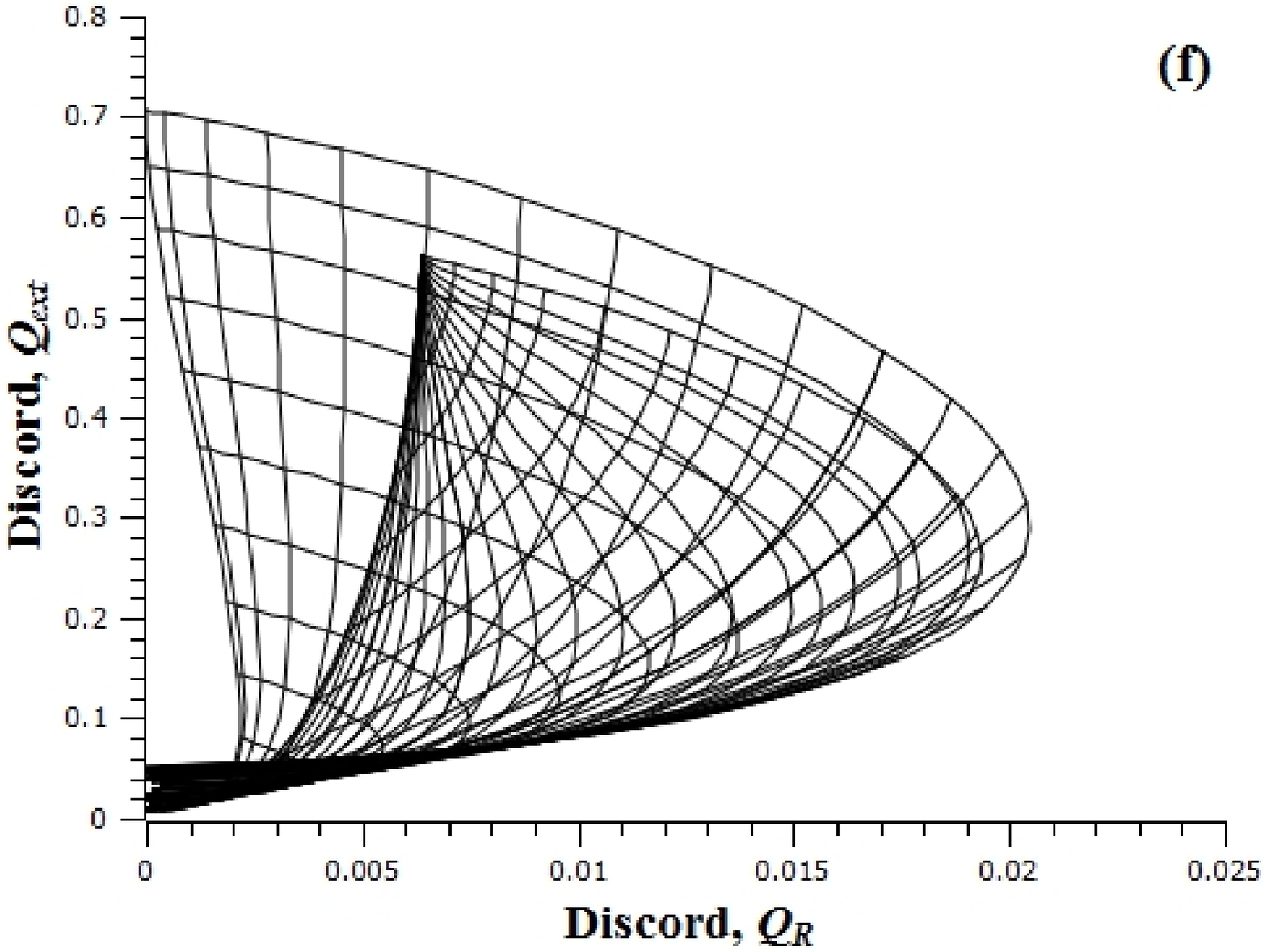,
  scale=0.35
   ,angle=0
}

\caption{The same as in Fig.\ref{Fig:Num200} for the chain of $N=200$ nodes.
(a,b,c) The parameters $\alpha_i$, $i=1,2$, 
vary 
inside of  sub-domain (\ref{reg}); (d,e,f) 
The parameters $\alpha_i$, $i=1,2$, vary 
inside of the whole domain (\ref{alpint}). (a,d) $\phi=\frac{3}{8}$, $t_0=55.91$, $R^2_{max}=0.97$; 
(b,e) $\phi=\frac{1}{4}$,
$t_0=69.48$, $R^2_{max}=0.88$;
(c,f) $\phi=0$ (homogeneous chain), $t_0=205.54$, $R^2_{max}=0.19$.
} 
  \label{Fig:Num200} 
\end{figure*}

 Decreasing the parameter $\phi$ from $\frac{1}{2}$ to 0 we slowly transform the Ekert chain to the 
 homogeneous one. 
 The results of the numerical  simulation of map (\ref{map})  for  chains of $20$ and $200$ spins and 
 $\phi=\frac{3}{8},\;\frac{1}{4},\;0$ are collected in Figs.\ref{Fig:Num20} and \ref{Fig:Num200}.
 As was mentioned above, the area of creatable region is minimal in the case of homogeneous chain $\phi=0$, 
 see Fig.\ref{Fig:Num20}(c,f) and   Fig.\ref{Fig:Num200}(c,f).

 In Figs.\ref{Fig:Num20}a-b and \ref{Fig:Num200}a-b, we depict
  map (\ref{map})
 corresponding to  sub-domain (\ref{reg}) of the  control-parameter space, 
 this is almost the one-to-one map. 
 We see that the  map in this case can be viewed as a deformation of the Ekert case shown in Fig.\ref{Fig:Ekert}a.
 Especially this is valid for $\phi=\frac{3}{8}$, see Figs.\ref{Fig:Num20}a and \ref{Fig:Num200}a. 
In addition, for $\phi=\frac{3}{8}$, we can partially realize the case of large $Q_R$ and small $Q_{exp}$
(the right lower  corners  in these figures correspond to the right upper corner in Fig.\ref{Fig:QQ}). 
 
Map (\ref{map}) of the whole domain  of control parameters (\ref{alpint}) into the creatable region
 is depicted in 
 Figs.\ref{Fig:Num20}d-f and \ref{Fig:Num200}d-f for the chains of 20 and 200 spins respectively. 
 In this case the map is far from 
 the one-to-one map with many mutual crossing of the 
 lines inside of the families $\alpha_1=const$ and $\alpha_2=const$. 
 We also note  that there is a sub-region near the ordinate axis (small $Q_R$) which is  
 covered four times by the parameters from each sub-domain (\ref{reg}-\ref{reg22}). Similar to the Ekert case, 
 the states from this sub-region are simpler creatable then states from the other sub-regions covered 
 tree-, two- and one-time.

\section{Conclusions}
\label{Section:conclusion}

In this work we study the possibility of remote creation of quantum correlated states.
We consider the model with the nearest neighbor XY-Hamiltonian 
and the coupling  constants depending on the parameter $\phi$ characterizing the inhomogeneity of the chain.
At that,   the
homogeneous chain corresponds to $\phi=0$, while the Ekert chain corresponds to
$\phi=\frac{1}{2}$. We consider the two parameters characterizing quantum correlations. The first one, $Q_{ext}$,
is  the discord between 
the receiver  and the rest of communication line, it shows whether the receiver is independent on 
other spins of the chain.
The second parameter, $Q_R$, is the discord between 
the two nodes of the receiver and characterizes   correlations inside of the receiver. 
We show that the creatable region increases with an increase in the  parameter $\phi$,
so that the  complete state-space in terms of the parameters $Q_{ext},Q_R$  can be 
covered in the case $\phi=\frac{1}{2}$ (Ekert chain). With decrease in $\phi$, the creatable
region reduces covering  the minimal area at $\phi=0$ (homogeneous chain). If $\phi=\frac{3}{8}$ 
(i.e. the almost Ekert chain),
the creatable region 
does not significantly reduces as can be seen in Figs.\ref{Fig:Num20}a,d and \ref{Fig:Num200}a,d. Moreover, 
comparing  Fig.\ref{Fig:Num20}a  with Fig.\ref{Fig:Num200}a allows us to conclude that, 
in this case, the area of creatable region slightly depends on $N$, which agrees with the
prediction of Sec.\ref{Section:R}. However, decreasing  $\phi$, we observe that the area of  creatable region
reduces with increase in $N$, which is especially evident from the comparison of
Fig.\ref{Fig:Num20}c(f) with Fig.\ref{Fig:Num200}c(f). The most interesting case corresponds to the right upper corner in 
Fig.\ref{Fig:QQ}a, where $Q_R$ reaches large values while $Q_{ext}$  is significantly less. 
In this case the quantum correlations between the receiver and the rest spins  of communication line  are minimal,
so that the receiver can be considered as an independent subsystem. 
The states from this corner can be created in the chains engineered for the high probability state transfer 
(see the right lower corners in  Fig.\ref{Fig:Ekert} (Ekert chain) and  in 
Figs.\ref{Fig:Num20}a,d and \ref{Fig:Num200}a,d, where  $\phi=\frac{3}{8}$).

We also emphasize that 
there is a domain in the control parameter space (\ref{reg}) which almost uniquely 
covers a large part of the creatable region, as shown in Figs.\ref{Fig:Num20}a-c and \ref{Fig:Num200}a-c. 
Outside of this domain the map loses its uniqueness, see  Figs.\ref{Fig:Num20}d-f and \ref{Fig:Num200}d-f.
In these figures, we also see that the subregion  near the ordinate axis ($Q_R$ is relatively small) 
is covered four times by the control parameters 
and thus it is simpler for realization in comparison with other subregions.

It is interesting that the inhomogeneity in our model establishes the lower limit on the state-transfer 
probability $R^2$ which  is estimated by the 
empirically obtained dash-curve in Fig.\ref{Fig:destr}.

This work is partially supported by the program of RAS 
''Element base of quantum computers'', project 
''Quantum registers on the virtual particles (fermions) in one-dimensional 
chains of interacting nuclear spins in the external magnetic 
field'',  by the Russian Foundation for Basic Research, grants No.15-07-07928. A.I.Z. is 
 partially supported by DAAD (the Funding program 
 ''Research Stays for University Academics and Scientists'', 2015
(50015559)).

\section{Appendix. Discord between two nodes of receiver}
\label{Section:appendix}


We call $Q_{N-1}$ and $Q_{N}$  discords calculated using the measurements over the  $(N-1)$th and $N$th  nodes respectively. 
First, we obtain the formula 
for $Q_N$:
\begin{eqnarray}\label{Q}
Q_N={\cal{I}}(\rho) -{\cal{C}}^N (\rho).
\end{eqnarray}  
Here   ${\cal{I}}(\rho)$ is the total mutual information \cite{OZ} which may be
written as follows: 
\begin{eqnarray}\label{I}
&&
{\cal{I}}(\rho) =S(\rho^{(N-1)}) + S(\rho^{(N)}) + \sum_{j=0}^1 \lambda_j \log_2
\lambda_j,\\\nonumber
\end{eqnarray}
where  $\lambda_j$ ($j=0,1$) are  the non-zero eigenvalues of the density
matrix $\rho^{R}$ (\ref{rhoB}),
\begin{eqnarray}
\lambda_0= \rho_{NN}+\rho_{(N-1)(N-1)},\;\;\lambda_1=1-\lambda_0.
\end{eqnarray}
Here $\rho_{ij}$ are the elements of the matrix $\rho^R$ ($\rho_{NN}=|f_N|^2$, $\rho_{(N-1)(N-1)}=|f_{N-1}|^2$), 
$\rho^{(N-1)}={\mbox{Tr}}_N \rho^{R}$ and $\rho^{(N)}={\mbox{Tr}}_{N-1}
\rho^{R}$ are the reduced density matrices,
 the entropies $S(\rho^{(N-1)})$ and $S(\rho^{(N)})$ are given by
the following formulas:
\begin{eqnarray}\label{SAB}
&&S(\rho^{(N-1)})=-(1- \rho_{NN} ) \log_2(1-\rho_{NN}) -
          \rho_{NN} \log_2\rho_{NN} ,\\\nonumber
&&S(\rho^{(N)})=-(1-\rho_{(N-1)(N-1)} ) \log_2(1-\rho_{(N-1)(N-1)}) -
 \rho_{(N-1)(N-1)} \log_2\rho_{(N-1)(N-1)} .
\end{eqnarray}
The so-called classical counterpart ${\cal{C}}^B (\rho^{R})$ of the mutual
information 
can be found considering the minimization over the projective measurements performed 
over
the $N$th spin \cite{ARA}:
\begin{eqnarray}\label{CB2}
&&
{\cal{C}}^{(N)} (\rho)=S(\rho^{(N-1)}) -\min\limits_{\eta\in[0,1]}(p_0 S_0 + p_1
S_1),
\end{eqnarray}
where
\begin{eqnarray}\label{S}
&&S(\theta_i)\equiv S_i = -\frac{1-\theta_i}{2}\log_2\frac{1-\theta_i}{2}-
                 \frac{1+\theta_i}{2}\log_2\frac{1+\theta_i}{2},
\\\label{p}
&&p_i=\frac{1}{2} \Big(1+(-1)^i\eta(1-2\rho_{(N-1)(N-1)}) \Big),\\\label{theta}
&&\theta_i=\frac{1}{p_i}\Big[(1-\eta^2) \rho_{(N-1)(N-1)}  \rho_{NN} +\\\nonumber
&&
\frac{1}{4}
\Big(
1-2\rho_{NN} +(-1)^i \eta(1-2(\rho_{(N-1)(N-1)}+\rho_{NN}))\Big)^2 
 \Big]^{1/2},
\\\nonumber
&&
i=0,1.
\end{eqnarray}
Here we introduce the parameter $\eta$ instead of $k$ in \cite{ARA}
($k=(1+\eta)/2$).
It is simple to show that the quantum discord $Q_{N-1}$ obtained performing 
 the von Neumann type measurements  on the particle $N-1$ can be calculated 
as follows:
\begin{eqnarray}\label{QA}
Q_{N-1}=Q_N|_{\rho_{(N-1)(N-1)}\leftrightarrow \rho_{NN}}.
\end{eqnarray}
Then we define the  discord $Q_{R}$ as the minimum of $Q_{N-1}$ and $Q_{N}$
\cite{FZ}, see eq.(\ref{final_discord_ex}).
%
One can show \cite{FZ_QIP2014} that the minimum in eq.(\ref{CB2})
corresponds to $\eta=0$  so that 
we result in the  explicit formulas (\ref{final_discord_ex}) (\ref{final_discord_ex0}) for the discord $Q_R$.

\begin{thebibliography}{99}

 
\bibitem{PBGWK2}
N.A.Peters, J.T.Barreiro,  M.E.Goggin, T.-C.Wei,  and P.G.Kwiat, Phys.Rev.Lett. {\bf   94}, 
150502 (2005) 

\bibitem{PBGWK}
N.A.Peters, J.T.Barreiro, M.E.Goggin, T.-C.Wei, and P.G.Kwiat in {\it Quantum
Communications and Quantum Imaging III}, ed. R.E.Meyers,
Ya.Shih, Proc. of SPIE {\bf   5893} (SPIE, Bellingham, WA, 2005) 

\bibitem{DLMRKBPVZBW}
B.Dakic, Ya.O.Lipp, X.Ma, M.Ringbauer, S.Kropatschek,
S.Barz, T.Paterek, V.Vedral, A.Zeilinger, C.Brukner, and P.Walther, 
Nat. Phys. {\bf   8}, 666 (2012). 

\bibitem{XLYG}
G.Y. Xiang, J.Li, B.Yu, and G.C.Guo
Phys. Rev. A {\bf   72}, 012315  (2005)

\bibitem{PSB}
S.Pouyandeh,  F. Shahbazi,  A. Bayat,
Phys.Rev.A {\bf 90}, 012337 (2014)

 
 
\bibitem{BBVB}
L.Banchi,  A. Bayat,  P. Verrucchi, and S.Bose, 
Phys.Rev.Let. {\bf 106}, 140501 (2011)
 
 
\bibitem{CS}
 B. Chen and Zh. Song, Sci. China-Phys., Mech.Astron. 53, 1266
(2010).
 
 
\bibitem{DSC}
A.Datta, A.Shaji,  C.M.Caves,
Phys.Rev.Lett. {\bf 100}, 050502 (2008)

\bibitem{LBAW}
 B.P.Lanyon, M.Barbieri, M.P.Almeida, A.G.White,
 Phys.Rev.Lett. {\bf 101}, 200501 (2008)

 
 
\bibitem{NLLZ}
W.J.Nie, Yu.H.Lan, Yo.Li, and Sh.Ya.Zhu, 
Sci.China-Phys., Mech. Astron {\bf 57}, 2276 (2014)

\bibitem{ZC}
P. Zhang, B. You, and L.-X. Cen,
Chin. Sci. Bull., {\bf 59}, 3841 (2014)

\bibitem{SXSZDWHCKW}
J.X. Sci, W. Xu, G. Sun et al,  Chin. Sci. Bull. {\bf 59}, 2547 (2014)

 \bibitem{RDL}
 S. Rodriques, N. Datta, and P. J. Love,
 Phys. Rev. A {\bf 90}, 012340 (2014) 

 
\bibitem{FBE}
E.B.Fel'dman, R.Br\"uschweiler and R.R.Ernst, Chem.Phys.Lett. {\bf 294}, 297
(1998)
 
 
 
\bibitem{Bose}
S. Bose, Phys. Rev. Lett. {\bf   91}, 207901 (2003)

 
\bibitem{BBCJPW}
C.H.Bennett, G.Brassard, C.Cr\'epeau, R.Jozsa, A.Peres, and W.K.Wootters,
Phys. Rev. Lett. {\bf   70}, 1895 (1993)


\bibitem{BPMEWZ}
D.Bouwmeester, J.-W. Pan, K.Mattle, M.Eibl, H.Weinfurter, and  A. Zeilinger, 
Nature {\bf 390}, 575 (1997)

\bibitem{BBMHP}
D. Boschi,  S. Branca,  F. De Martini, L. Hardy,  and S. Popescu,
Phys. Rev. Lett. {\bf 80}, 1121 (1998)


 
\bibitem{CDEL}
 M.Christandl, N.Datta, A.Ekert and A.J.Landahl, Phys.Rev.Lett. {\bf   92}, 187902 (2004)


\bibitem{ACDE}
 C.Albanese, M.Christandl, N.Datta and A.Ekert, Phys.Rev.Lett. {\bf   93}, 230502 (2004)

\bibitem{KS}
 P.Karbach and J.Stolze, Phys.Rev.A {\bf   72}, 030301(R) (2005)

 
\bibitem{GKMT}
 G.Gualdi, V.Kostak, I.Marzoli and P.Tombesi, Phys.Rev. A {\bf   78}, 022325 (2008)



\bibitem{BDSSBW}
C.H.Bennett, D.P.DiVincenzo, P.W.Shor, J.A.Smolin, B.M.Terhal, and W.K.Wootters,
Phys.Rev.Lett. {\bf   87}, 077902 (2001);
 Erratum,
 C.H.Bennett, D.P.DiVincenzo, P.W.Shor, J.A.Smolin, B.M.Terhal, and W.K.Wootters, 
 Phys. Rev. Lett. {bf   88}, 099902(E) (2002)
 
 
\bibitem{BHLSW}
C.H.Bennett, P.Hayden,
D.W.Leung, P.W.Shor, and A.Winter, 
IEEE Transetction on Information Theory {\bf   51}, 56 (2005)  

 
 
\bibitem{G}
G.L.Giorgi, Phys. Rev. A {\bf   88}, 022315 (2013) 

 
 
\bibitem{KZ_2008}
 E.I.Kuznetsova and A.I.Zenchuk, Phys.Lett.A {\bf   372},  pp.6134-6140 (2008)

 

 \bibitem{SAOZ}
 J.Stolze, G. A. \'Alvarez,
O. Osenda, A. Zwick in
{\it Quantum State Transfer and Network Engineering.
Quantum Science and Technology},
ed. by  G.M.Nikolopoulos and I.Jex, Springer Berlin Heidelberg, Berlin, p.149  (2014) 


\bibitem{BK}
A. Bayat and V. Karimipour 
Phys.Rev.A {\bf 71}, 042330 (2005)

\bibitem{C}
P. Cappellaro,
Phys.Rev.A {\bf 83}, 032304 (2011)

\bibitem{QWL}
W. Qin,  Ch. Wang,   G. L. Long, 
Phys.Rev.A {\bf 87}, 012339 (2013)

 
 
\bibitem{B}
A.Bayat,
Phys. Rev. A {\bf 89}, 062302 (2014)


 
\bibitem{JKSS}
C. Godsil, S. Kirkland, S. Severini,  Ja. Smith 
Phys. Rev. Lett. {\bf 109}, 050502 (2012)

\bibitem{SO}
R.Sousa, Ya. Omar,
New J. Phys. {\bf 16}, 123003 (2014).

 
 
 \bibitem{BB}
 D. Burgarth and S. Bose,
Phys.Rev.A {\bf 71}, 052315 (2005)

\bibitem{SJBB}
K. Shizume,  K. Jacobs,  D. Burgarth,  and S. Bose,
Phys. Rev. A {\bf 75}, 062328 (2007)
 
 
\bibitem{Z_2014}
A.I.Zenchuk, 
Phys. Rev. A {\bf 90}, 052302(13) (2014) 

 
\bibitem{BZ_2015}
G. A. Bochkin and A. I. Zenchuk, 
Phys.Rev.A 91, 062326(11) (2015)
 
 \bibitem{FKZ_LANL}
  E.B. Fel'dman, E.I. Kuznetsova, A.I. Zenchuk,
 arXiv:1507.07738 

\bibitem{YBB}
S.Yang,  A. Bayat, S. Bose,
Phys.Rev.A {\bf 84}, 020302 (2011)


 
\bibitem{Z_2012}
A.I.Zenchuk,
J. Phys. A: Math. Theor. {\bf   45} (2012) 115306
 
\bibitem{PS}
S. Pouyandeh, F. Shahbazi, 
Int. J. Quantum Inform. {\bf 13}, 1550030 (2015)



\bibitem{HV}
L. Henderson, V. Vedral, J. Phys. A: Math. Gen. {\bf   34}, 6899 (2001)

\bibitem{OZ}
H.Ollivier and W.H.Zurek, Phys.Rev.Lett. {\bf 88}, 017901 (2001) 


\bibitem{Zurek}
W.H. Zurek, Rev. Mod. Phys. {\bf   75}, 715 (2003)


\bibitem{Wootters}
 W.K. Wootters,,
Phys. Rev. Lett. {\bf   80},
2245 (1998)

\bibitem{HW}
S.Hill and W.K.Wootters, Phys. Rev. Lett. {\bf    78}, 5022 (1997)



\bibitem{P}
A.Peres, Phys. Rev. Lett. {\bf    77}, 1413 (1996)


\bibitem{AFOV}
L.Amico, R.Fazio, A.Osterloh and V.Ventral, Rev. Mod. Phys. {\bf    80}, 517 
(2008)

\bibitem{HHHH}
R.Horodecki,
P.Horodecki, M.Horodecki and K.Horodecki, Rev. Mod. Phys. {\bf    81}, 865  (2009)


 \bibitem{FZ_QIP2014}
 E. B. Fel'dman and A. I. Zenchuk,
 Quantum Inf. Process., {\bf 13} (2014) 201
 
 
\bibitem{ARA} 
 M.Ali, A.R.P.Rau, G.Alber, Phys.Rev.A {\bf 81}, 042105 (2010); Erratum: 
 Phys.Rev.A {\bf 82}, 069902(E) (2010)

 
\bibitem{FZ}
E.B.Fel'dman and A.I.Zenchuk,
Phys. Lett. A {\bf   373} (2009) 1719
 
 
 

\end{thebibliography}
\end{document}